\documentclass[fleqn,usenatbib]{mnras}
\usepackage{amsmath}
\usepackage{newtxtext,newtxmath}

\usepackage[T1]{fontenc}
\usepackage{comment}
\DeclareRobustCommand{\VAN}[3]{#2}
\let\VANthebibliography\thebibliography
\def\thebibliography{\DeclareRobustCommand{\VAN}[3]{##3}\VANthebibliography}

\usepackage{graphicx}	
\usepackage{amsmath}	

\title[Optical and X-ray studies of 1A~0535+262]{Optical and X-ray studies of Be/X-ray binary 1A~0535+262 during its 2020 giant outburst}

\author[Chhotaray et al.]{
Birendra Chhotaray$^{1,2}$\thanks{E-mail: birendra@prl.res.in, rsbirendra786@gmail.com},
Gaurava K. Jaisawal$^{3}$,
Neeraj Kumari$^{1}$,
Sachindra Naik$^{1}$,
Vipin Kumar$^{1}$,
\newauthor 
and Arghajit Jana$^{4}$
\\
$^{1}$Astronomy and Astrophysics Division, Physical Research Laboratory, Navrangpura, Ahmedabad - 380009, Gujarat, India\\
$^{2}$Indian Institute of Technology Gandhinagar, Palaj, Gandhinagar - 382055, Gujarat, India\\
$^{3}$ DTU Space, Technical University of Denmark, Elektrovej 327-328, DK-2800 Lyngby, Denmark\\
$^{4}$Institute of Astronomy, National Tsing Hua University, Hsinchu, 30013, Taiwan}
\date{Accepted XXX. Received YYY; in original form ZZZ}

\pubyear{2022}

\begin{document}
\label{firstpage}
\pagerange{\pageref{firstpage}--\pageref{lastpage}}
\maketitle

\begin{abstract}
We report results obtained from the optical and X-ray studies of the Be/X-ray binary 1A~0535+262/HD~245770 during the 2020 October giant X-ray outburst, using the 1.2 m telescope at Mount Abu Infrared observatory and AstroSat, respectively. The peak flux of the outburst was recorded to be $\sim$11 Crab in the 15-50 keV range, the highest ever observed from the pulsar. We performed optical observations in the 6000-7200~\AA~band before, during, and after the outburst to investigate the evolution of the circumstellar disc of the Be star between 2020 February and 2022 February. Our optical spectra exhibit prominent emission lines at 6563~\AA (H~I), 6678~\AA (He~I), and 7065~\AA (He~I). We found a significantly variable H$\alpha$ line in the spectra. The single-peaked line profile appeared asymmetric with broad red- \& blue-wings in the data before and during the outburst. The post-outburst observations, however, resulted in a double-peaked  profile with asymmetry in the blue-wing. Our observations before the outburst confirmed a larger Be disc that decreased in size as the outburst progressed. Furthermore, the observed variabilities in the H$\alpha$ line profile and parameters suggest the presence of a highly misaligned, precessing, and warped Be disc. AstroSat observation of the pulsar detected pulsations at $\sim$103.55 s in the light curve up to 110 keV. We found strongly energy-dependent pulse profiles with increasing contribution of the pulsing component in hard X-rays. The broadband spectral fitting in the 0.7-90.0 keV range confirmed the presence of the known cyclotron resonance scattering feature at $\sim$46.3 keV.
\end{abstract}




\section{Introduction}

Be/X-ray binaries (BeXRBs) represent the largest population of high mass X-ray binaries (HMXBs). These BeXRBs primarily consist of a neutron star as the compact object and a massive (10-20 M$_{\odot}$) non-supergiant star as the optical companion rotating around the common center of mass (e.g.,~ \citealt{ziolkowski2002x, paul2011transient, reig2011x}). The neutron stars in these binary systems are powered by mass accretion from the companion Be star. In BeXRBs, the companion Be star is bright in optical and infrared wavebands. It is an early-B or late-O type star that commonly shows H~I, He~I, and Fe~II emission lines at a specific phase of their lifetime \citep{porter2003classical}. The Be stars also show an excess in the infrared domain of the electromagnetic spectrum. The emission lines in the optical and infrared spectra and the infrared excess over the thermal continuum are unique characteristics of the Be stars, unlike the classical O \& B-type stars where the photospheric absorption lines are seen in the optical/infrared spectra without any signature of the infrared excess. Such distinctive characteristics are attributed to the presence of an equatorial circumstellar disc around the Be star. The circumstellar disc is understood to be formed due to the discharge of material from the fast rotating Be star, close to the critical velocity \citep{porter2003classical}.

The orbit of  most of the BeXRBs is wide and eccentric (e~$\geq$~0.3) with an orbital period of $\geq$10 d. The mass transfer from the Be star to the compact object leads to a transient X-ray outburst at the periastron passage. However, all BeXRBs are not transient in nature \citep{1999MNRAS.306..100R}. The BeXRBs show two types of X-ray activities: normal (Type~I) and giant (Type~II) X-ray outbursts. The normal outbursts are periodic or quasi-periodic events that occur close to the periastron passage of the neutron star. These outbursts last for a fraction ($\sim$20\%) of the orbital period with a peak luminosity reaching up to $L_{\rm X}$ $\sim$ 10$^{35-37}$ erg s$^{-1}$. On the other hand, the giant X-ray outbursts are irregular and rare. They do not show any orbital modulation and last for several orbital periods with a peak luminosity of $L_{\rm X}$ $\geq$10$^{37}$ erg s$^{-1}$. It is now widely accepted that the mass transfer from the Be circumstellar disc to the neutron star commences X-ray outbursts, though the exact mechanism behind this is still a matter of study.

\citet{okazaki2001natural} suggested that normal X-ray outbursts occur in the BeXRBs with medium to high eccentric orbits and can be explained using the resonantly truncated decretion disc model \citep{2001A&A...369..108N,10.1046/j.1365-8711.2002.05960.x}. \citet{10.1093/pasj/65.2.41} presented a detailed scenario of accretion processes using two of the most studied systems such as 1A~0535+262 and 4U~0115+634. Based on their study, the normal outbursts seen in the BeXRBs are caused by the radiatively inefficient accretion flows (RIAFs) from a tidally truncated Be disc to the neutron star. They also suggest that the giant X-ray outbursts occur in the systems where the disc of the Be star is misaligned with the binary orbital plane. In such cases, it is proposed that the outer part of the Be disc is warped (e.g., \citealt{1999A&A...348..512P} \& \citealt{10.1111/j.1365-2966.2011.19231.x}), and the neutron star accretes matter via Bondi-Hoyle-Lyttleton (BHL) accretion. \citet{Martin_2014_a} suggested that the giant X-ray outbursts originate in a highly misaligned, eccentric, and warped Be disc.

The warping episode of a disc was also suggested for several systems such as $\gamma$~Cas \& 59~Cyg \citep{1998A&A...330..243H}, 4U~0115+63 \citep{2007A&A...462.1081R,2018A&A...619A..19R}, 1A~0535+262 \citep{article_Moritani} through observations. \citet{1998A&A...330..243H} proposed an inclined Be disc with respect to the orbital plane by analysing the variation in emission lines from the shell profile to the single-peaked profile for $\gamma$~Cas \& 59~Cyg. From the long-term optical/IR photometric observations and optical spectroscopic observations of 4U~0115+63/V635~Cas,  \citet{2007A&A...462.1081R} interpreted the presence of a warped Be disc at the time of the giant outburst. Later, \citet{2018A&A...619A..19R} also presented the evidence of a warped disc for 4U~0115+63/V635~Cas from the polarimetric and spectroscopic variability studies during the 2015 and 2017 giant outbursts. 

1A~0535+262 is one of the most active BeXRB transients with frequent Type~I X-ray outbursts. The Ariel~V space telescope discovered this source in 1975 during a giant X-ray outburst  \citep{rosenberg1975observations}. The system consists of a neutron star orbiting around an O9.7~IIIe star \citep{giangrande1980optical} with an orbital period ($P_{\rm orb}$) of $\sim$110 days in a relatively wide and eccentric orbit (eccentricity ($e$) $\sim$0.47) \citep{finger1994hard}. The spin period of the neutron star was estimated to be $\sim$104 s \citep{rosenberg1975observations}. 1A~0535+262 underwent several giant outbursts since its discovery (\citet{Camero_Arranz_2012} \& references therein). \citet{10.1046/j.1365-8711.1999.02112.x} found that the X-ray outburst in 1A~0535+262 occurred during the optical fading phase of the Be star. \cite{Camero_Arranz_2012} used the contemporaneous X-ray and optical data spanning over thirty years and got an anti-correlation between the $V$-band magnitude and the equivalent width of the H$\alpha$ line before giant outbursts. \citet{Yan_2011} interpreted the observed anti-correlation between the $V$-band magnitude and H$\alpha$ line equivalent width before the 2009 giant X-ray outburst in terms of the mass ejection process from the inner part of the circumstellar disc. From spectroscopic observations, a strong H$\alpha$ line variability was observed in 1A~0535+262/V725~Tau \citep{10.1093/pasj/63.4.L25}. Before the giant outburst, the H$\alpha$ line is observed to be evolving from any shape to single peak with increasing strength and Full width at Zero Intensity (FWZI) \citep{1998MNRAS.294..165C,10.1111/j.1365-2966.2006.10127.x, Camero_Arranz_2012}. This behaviour of H$\alpha$ line is attributed to the warping of Be companion circumstellar disc before the giant outburst. Using the high dispersion optical spectroscopic observations during the 2009 and 2011 giant X-ray outbursts, \citet{article_Moritani} suggested that a precessing warped Be disc triggered the giant outburst in 1A~0535+262. 

While the optical and infrared observations of the BeXRBs provide information on the properties of the Be star and the circumstellar disc around it, the observations in X-ray domain can be used to understand the effect of mass transfer onto the neutron star and its ambient environment. Timing analysis of the data obtained during several giant outbursts of 1A~0535+262 showed that the pulsar spun-up with the increase in its X-ray luminosity \citep{1997ApJS..113..367B, Sartore_2015}. The spinning up of the pulsar is interpreted as due to the angular momentum transfer from the accretion flow to the neutron star which rises with the increase in the accretion rate. Change in the shape of the pulse profile of the pulsar with luminosity is understood to be due to the change in the accretion environment around the neutron star with the change in luminosity \citep{1989ApJ...338..373P}. During the 2011 giant outburst, this feature was also observed in 1A~0535+262 \citep{Sartore_2015}. During X-ray outbursts, the broad-band X-ray spectrum of 1A~0535+262 has been primarily modeled with a cutoff power-law continuum model along with a blackbody component of temperature in the 1-2 keV range for soft X-ray excess and a Gaussian function at 6.4 keV for iron emission line \citep{Caballero_2013,Sartore_2015}. The cyclotron resonance scattering features, the tools for the direct measurement of pulsar's magnetic field, with fundamental and first harmonic at $\sim$ 45 keV and $\sim$ 100 keV, respectively, are also observed in the spectrum of 1A~0535+262 \citep{1994A&A...291L..31K,2006ApJ...648L.139T,Sartore_2015,2021ApJ...917L..38K}. Though the change in the cyclotron line energy with luminosity in 1A~0535+262 does not show a clear trend, some of the measurements support a positive correlation \citep{refId013,Sartore_2015}. 

Recently, 1A~0535+262 underwent a giant outburst which was triggered on MJD 59151 \citep{2020ATel14157....1M}. The Swift/BAT monitoring data showed that the peak intensity of the pulsar during this outburst reached a level of $\sim$11 Crab in the 15-50 keV range on MJD 59172 (2020 November 19). This peak intensity is the highest among the previously observed giant X-ray outbursts of 1A~0535+262 \citep{Camero_Arranz_2012} since its discovery \citep{Jaisawal2020ATel14179....1J, Jaisawal2020ATel14227....1J}. To investigate the changes in the properties of the Be star and its circumstellar disc, we observed the source before, during, and after the giant X-ray outburst in the optical band using the 1.2 m telescope at Mount Abu Infrared Observatory (MIRO). Apart from the optical observations, we also carried out a pointed X-ray observation of the pulsar with AstroSat during the rising phase of the outburst to investigate its X-ray characteristics. The paper is organized as follows. In Section~2, we report the optical spectroscopic and X-ray observations of the system and corresponding data reduction procedure. The results obtained from the optical observations are presented in Section~3, followed by X-ray data analysis results in Section~4. In the end, Section~5 presents the discussion of our results and a summary of our findings. 

\section{Observations \& Data Reduction}
\subsection{Optical Observations}
X-ray outbursts in the BeXRB systems are known to be due to the abrupt accretion of a sufficiently large amount of matter from the Be circumstellar disc onto the neutron star. Considering this, the X-ray properties of the neutron star during outbursts must be associated with the change in the Be star properties. Hence, studying  such BeXRBs in optical/infrared and X-ray bands during X-ray outbursts is very important. To compare the properties of the Be star during X-ray outbursts, the optical/infrared observations of the Be star during pre- and post-outburst phases are useful. 

\begin{figure*}
    \hspace*{-1.2cm}
    \includegraphics[height=7cm, width=20cm]{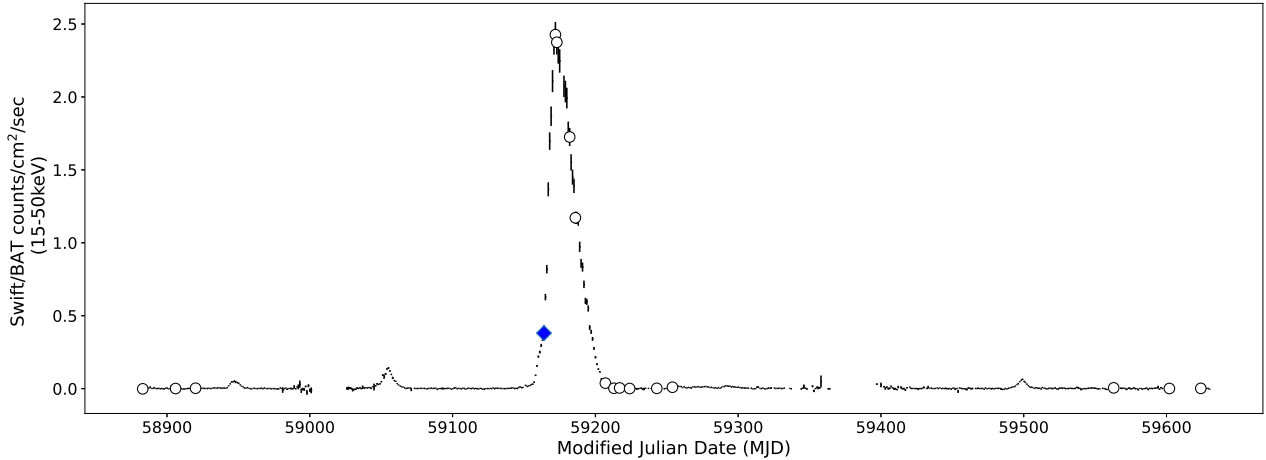}
    \caption{Swift/BAT daily light curve of the pulsar 1A~0535+262 in 15-50 keV range, covering the 2020 giant X-ray outburst. The black open circles represent the epochs of optical observations of the binary system with the 1.2 m telescope at Mount Abu using the MFOSC-P instrument. The blue diamond represents the epoch of AstroSat observation of the pulsar. The source was extremely bright with an intensity of $\sim$11 Crab at the X-ray outburst peak at around MJD 59172; for reference, 1~Crab represents 0.220 counts~cm$^{-2}$~s$^{-1}$ in the 15-50 keV band.}
    \label{fig:swiftlc}
\end{figure*}

As mentioned, the neutron star in the Be/X-ray binary system 1A~0535+262/HD~245770 went into a giant X-ray outburst in 2020 October. As part of the ongoing project, we had optical observations of the system at a few epochs before it went into the X-ray outburst. We continued our optical spectroscopic observations of 1A~0535+262 during the X-ray outburst using the 1.2 m telescope of the Mount Abu Infrared Observatory with the Faint Object Spectrograph and Camera-Pathfinder (MFOSC-P) instrument \citep{2018SPIE10702E..4IS,srivastava2021design} mounted on the 1.2 m, f/13 telescope at several epochs. The instrument is designed to provide seeing limited imaging in Bessel BVRI filters with a sampling rate of 3.3 pixels per arc-second over a 5.2$\times$5.2 arc-minute$^{2}$ field-of-view. The MFOSC-P was facilitated with three plane reflection gratings, named R2000, R1000 and R500 with 500, 300 and 150 lp/mm at wavelengths of $\sim$6500~\AA, 5500~\AA~ and 6000~\AA, respectively. R2000, R1000 and R500 represent the spectral resolutions (R=$\frac{\lambda}{\Delta\lambda}$) of 2000, 1000 and 500, respectively. These three modes provide a standard spectral coverage of $\sim$6000-7000~\AA, $\sim$4700-6650~\AA~ and $\sim$4500-8500~\AA, respectively. The spectroscopic observations were carried out on several nights (see Table~\ref{tab:log} \& Figure~\ref{fig:swiftlc}) using a 75 $\mu$m slit which is equivalent to 1$^{''}$ on the sky. In the present work, the spectroscopic observations were carried out in R2000 mode, the highest resolution available for the MFOSC-P instrument.

The raw data of the MFOSC-P observations were reduced using the in-house developed data analysis routines in Python\footnote{ https://www.python.org/}  with the assistance of open-source image processing libraries (ASTROPY\footnote{ https://www.astropy.org/}, etc.) (also refer to \citealt{2022MNRAS.510.4265K} for the analysis method). In the beginning, the bias subtracted, cosmic ray removed and sky subtracted images were created. Using a halogen lamp, the pixel-to-pixel efficiency variation was found to be less than 1\% and hence no correction was applied. In the second step, the spectra of Neon and Xenon calibration lamps were taken after each exposure for the wavelength calibration of the data. For the instrument response correction, a nearby field star SAO~77466, observed at a similar airmass as HD~245770, was used as the standard star for spectroscopic observations. The response function is calculated by dividing the observed continuum spectra of SAO~77466 with a blackbody curve corresponding to the standard star's effective temperature. Finally, the science images were created by applying the response function to the source images.


\subsection{X-ray Observation}
AstroSat is India's first multi-wavelength astronomical satellite \citep{agrawal2006broad} that is dedicated for the observations of cosmic sources in a broad energy range starting from optical to X-ray bands. It was launched by the Indian Space Research Organization on 2015 September 28. AstroSat is equipped with five major payloads: Ultra Violet Imaging Telescope (UVIT; \citealt{2017AJ....154..128T}), Scanning Sky Monitor (SSM; \citealt{ramadevi2017early}), Soft X-ray Telescope (SXT; \citealt{2017JApA...38...29S}), Large Area X-ray Proportional Counter (LAXPC; \citealt{yadav2016large}), and Cadmium Zinc Telluride Imager (CZTI; \citealt{bhalerao2017cadmium}). Along with the  payloads mentioned above, an auxiliary payload known as Charged Particle Monitor (CPM) is also deployed to apprise other detectors about the presence of the South Atlantic Anomaly (SAA) region along the orbit.

In the present work, we use data from a Target of Opportunity (ToO) observation of 1A~0535+262 on 2020 November 11 (MJD 59165), in the rising phase of its giant X-ray outburst (Figure~\ref{fig:swiftlc}). The details of the observation are given in Table~\ref{tab:log}. During the ToO observation, the UVIT was not operational. The SXT is a soft X-ray focusing telescope and is sensitive to photons in the 0.3-8 keV range. Due to its focusing characteristics, it is close to 1000 times more sensitive than any other non-focusing instruments with the same photon collecting area. The on-axis effective area of SXT is 90 cm$^{2}$ at 1.5 keV with an energy resolution of 90 eV and 136 eV at 1.5 keV and 5.9 keV, respectively. The time resolution of SXT is 2.4 s in \textit{Photon Counting (PC)} mode. The SXT transmits the recorded observed data after each orbit to the ground station. The level2 data from the 17 orbits of the PC mode observation are downloaded from the \textit{AstroSat Archive-ISSDC}. For further analysis, a single \textit{$event file$} was created by combing the event files of all orbits using the \textit{sxteventmergertool} module. Using the \textit{FTOOLS} task \textit{XSELECT} and considering a  circular region of size $15^{'}$ from the center of the Point Spread Function (PSF), the science products were generated from the merged event file. In SXT PC mode, the observations of sources fainter than $\sim$200 mCrab do not get affected by photon pile-up. As the intensity of 1A~0535+262 was more than 1~Crab during our AstroSat observation, the SXT data were severely affected due to photon pile-up. The data were corrected for piled-up effect by excluding the central circular region of four arc-min radius while extracting light curves and spectra. Corresponding effective area file ({\it ARF}) was also generated. The SXT team has provided the  \textit{sxt\_pc\_mat\_g0.rmf} \& \textit{SkyBkg\_comb\_EL3p5\_Cl\_Rd16p0\_v01.pha} files that were used for spectral redistribution matrix and background spectrum, respectively. 

The LAXPC consists of three identical, co-aligned, independently operated proportional counters. These detectors are sensitive to X-ray photons with energy in the range of 3-80 keV and provide a total effective area of 8000 $\rm cm^{2}$ at 5-30 keV. The energy and time resolutions of the LAXPC units are 12\% at 22 keV and 10 $\rm \mu$s, respectively. The dead time of the detectors is 42 $\rm \mu$s. Data from the currently working LAXPC20 unit are used in the present work. Data from LAXPC20 were further processed using the latest \textit{LAXPCsoftware}. Standard procedures were followed to generate source and background light curves and spectra using the \textit{event-analysis} mode data.

The CZTI is designed to perform simultaneous X-ray imaging and spectroscopic observations in the energy range of 20-150 keV. It is a large semiconductor pixelated CdZnTe detector with other ancillaries like a coded mask and collimator. The CZTI has a total effective area of 2000 $\rm cm^{2}$ for the energy range of 20-300 keV. Time tagging of an event can be estimated within 20 $\mu$s. Screening of the data, cleaning, and high-level products are generated from level1 data using modules provided within the \textit{CZTpipeline} task.

In addition to the AstroSat, we used the X-ray monitoring data from Swift/BAT daily light curve in the 15-50 keV range \citep{Krimm2013ApJS..209...14K}. The BAT data were filtered to exclude rows marked as bad quality data by using DATA\_FLAG=0.

\begin{table}
	\centering
	\caption{Log of optical and X-ray observations of the BeXRB 1A~0535+262/HD~245770 }
	\label{tab:log}
	\begin{tabular}{lccc} 
		\hline

 Date of & MJD  &  Days since & Exposure    \\ 
 
 observation &  & outburst & time (s)  \\ 
 
            &  & (29 Oct 2020) &  \\ [4 pt]
\hline 

\large{\textit{MIRO}}\\ [4 pt]

04 FEB 2020 & 58883 & $-276$ & 1000  \\ [4pt]
27 FEB 2020 & 58906 & $-253 $ & 1000 \\ [4pt]
12 MAR 2020 & 58920 & $-239$  & 1000 \\[4pt]
19 NOV 2020 & 59172 & $+13$ & 400 \\ [4pt]
20 NOV 2020 & 59173 & $+14$ & 1000 \\[4pt]
29 NOV 2020 & 59182 & $+23$ & 400 \\ [4pt]
03 DEC 2020 & 59186 & $+27$ & 250  \\[4pt]
24 DEC 2020 & 59207 & $+48$  & 800 \\[4pt]
30 DEC 2020 & 59213 & $+54$ & 500  \\[4pt]
03 JAN 2021 & 59217 & $+58$ & 700 \\ [4pt]
10 JAN 2021 & 59224  & $+65$ & 500  \\[4pt]
29 JAN 2021 & 59243 & $+84$ & 500 \\[4pt]
09 FEB 2021 & 59254 & $+94$ & 1000 \\ [4pt]
15 DEC 2021 & 59563 & $+404$ & 800 \\[4pt]
23 JAN 2022 & 59602 & $+443$ & 350 \\ [4pt]
14 FEB 2022 & 59624 & $+465 $ & 600 \\ [4pt]
\hline
\large{\textit{AstroSat}}  \\ [4pt]
11 NOV 2020 & 59165 & $+6$ & 94331 \\ [4pt]
\hline
\end{tabular}
\end{table}

\section{Results from optical spectroscopy}

In BeXRBs, the emission lines in the optical/infrared spectrum and infrared excess are attributed to the circumstellar disc around the Be star. Therefore, the changes in the strength and profile of the emission lines in the optical spectrum directly relate to the change in the dynamics of the Be disc. In this section, we present the observed changes in emission line parameters of 1A~0535+262/HD~245770 during the 2020 giant X-ray outburst compared to that of before and after the event using our optical observations.

\begin{figure*}
    \includegraphics[height=22cm, width=18cm]{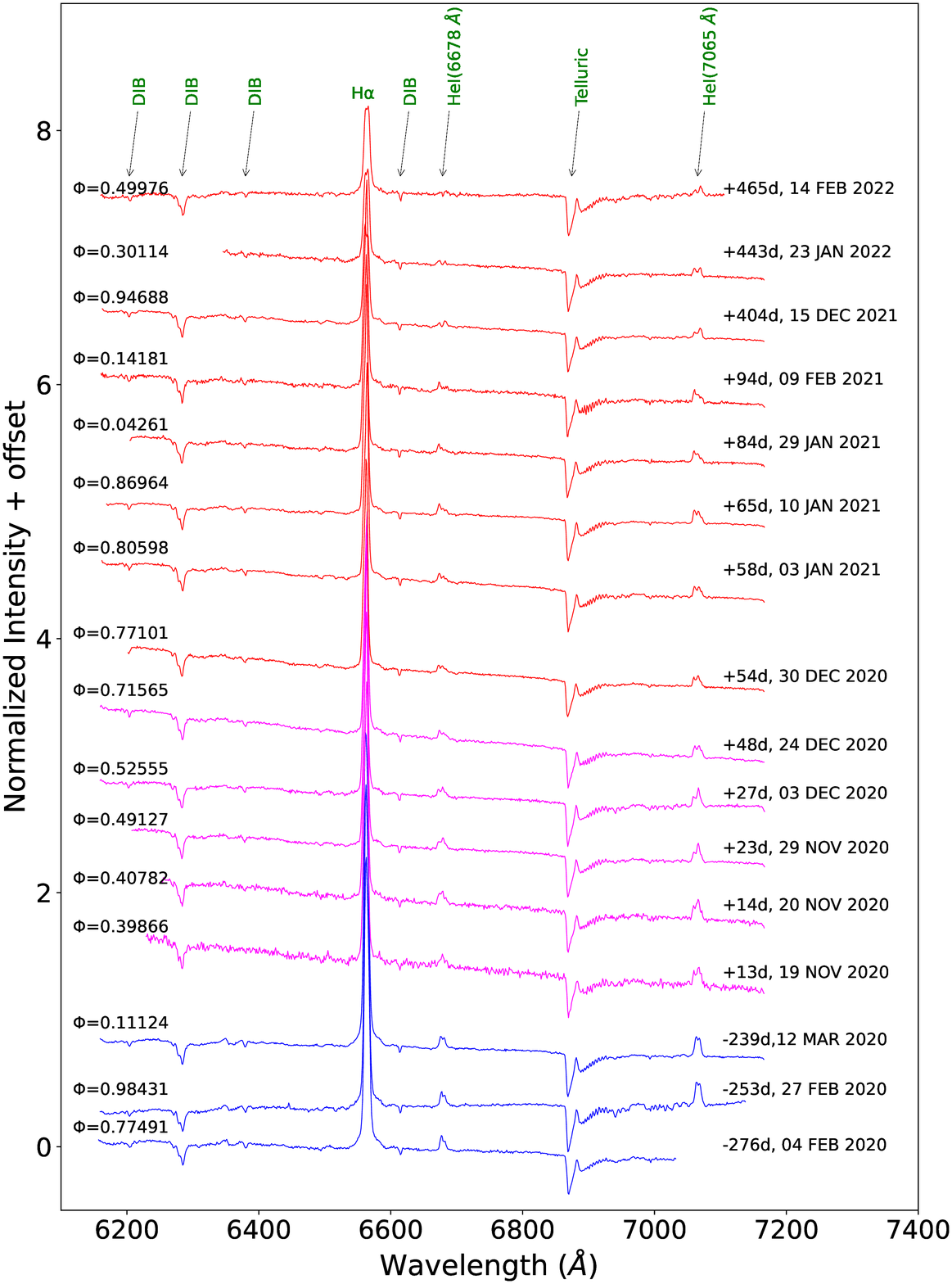}
    \caption{Evolution of the optical spectrum of Be/X-ray binary 1A 0535+262 as seen with the MFOSC-P instrument. The spectra are not reddening corrected. The observation date and orbital phase ($\phi$) of the neutron star are annotated on each spectrum. The numbers written in unit of days in minus and plus correspond to days spent before and after the onset of the giant X-ray outburst on MJD 59151, respectively. Blue, magenta, and  red colors represent the spectra obtained before, during, and after the giant outburst, respectively.}
    \label{fig:spec_evolve}
\end{figure*}

As mentioned above, the optical spectroscopic observations of the BeXRB were carried out in the 6000-7200 \AA \hspace{0.01cm} band using the MFOSC-P instrument in R2000 mode. Initial observations were made on 2020 February 4, February 27 and March 12, when the neutron star in the system was in the quiescent state (see Figure~\ref{fig:swiftlc}). These observations provide information on the Be circumstellar disc prior to the giant X-ray outburst. As the neutron star went into the giant X-ray outburst, we carried out closely spaced optical spectroscopic observations from 2020 November 19 (close to the peak of the X-ray outburst) to 2021 February 9 when the neutron star returned to the X-ray quiescent phase. The next three observations were made on 2021 December 15, 2022 January 23, and 2022 February 14 when the system was in the quiescent state. 

\subsection{ Optical spectra}

The 6000-7200 \AA \hspace{0.01cm} range spectra obtained from the MFOSC-P observations of 1A~0535+262/HD~245770 in R2000 mode are shown in Figure~\ref{fig:spec_evolve}. In the figure, the continuum normalized intensity for each epoch of observation is shown with offsets for clarity. The date of the corresponding observation and the number of days before and after the onset of the giant X-ray outburst are quoted in the figure for individual observations. The spectra in blue color were obtained before the giant outburst. The magenta and  red color spectra were obtained during and after the giant outburst, respectively. The numbers in the unit of days in minus and plus signs correspond to the epoch of the optical observation before and after the beginning of the outburst, respectively. The mentioned orbital phases ($\phi$) depict the position of the neutron star on its orbital path during the optical observation. The orbital phase is estimated using the method described by \citet{10.1111/j.1365-2966.2010.16454.x} and $\phi$=0 corresponds to the periastron passage of the neutron star.

The Diffuse Interstellar Bands (DIBs) and atmospheric telluric features were detected in the spectra from all the epochs of observations (Figure~\ref{fig:spec_evolve}). The DIBs are absorption features that arise when photons travel through significant column densities in the interstellar medium. Nevertheless, the DIBs are not well identified. Throughout our observations, four DIBs were detected at 6203.06 \AA, 6283.86 \AA,  6379.20 \AA, and 6613.62 \AA \hspace{0.1cm} \citep{1995ARA&A..33...19H} in the spectra. The telluric features arise because of the absorption of photons by the molecules present in the earth's atmosphere. These telluric features superimpose on the stellar spectra. One of the most prominent telluric features such as O$_{2}$ B band at $\lambda \sim 6870 $ \AA ~(\citealt{article} \& reference therein) was present in our spectra.

In the 6000-7200 \AA \hspace{0.01cm} range spectra of 1A~0535+262/HD~245770, spanned over two years of duration, three emission lines at 6562.8 \AA, 6678 \AA ~ and 7065 \AA \hspace{0.01cm} are detected (Figure~\ref{fig:spec_evolve}). The presence of emission lines in the optical spectra confirms the existence of the circumstellar disc in the BeXRB system \citep{10.1111/j.1365-2966.2006.10127.x}. This signifies that the disc is not entirely vanished even during/after the giant X-ray outburst (see, e.g.,   \citealt{10.1046/j.1365-8711.1999.03148.x} for 1A~0535+262 and \citealt{2014A&A...561A.137R} for IGR~J21343+4738).

\subsection{Evolution of emission line profile}

\begin{figure}
    \vspace{-1.3cm}
    \includegraphics[height=21.7cm,width=8cm]{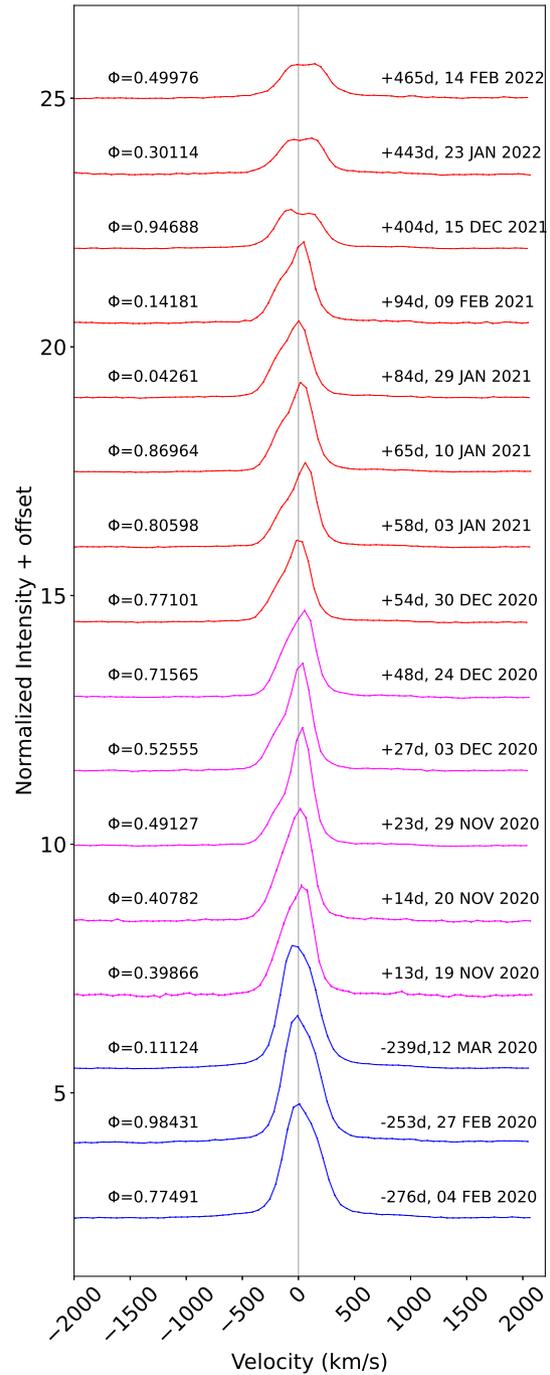}
    \caption{The H$\alpha$ line profiles of 1A~0535+262/HD~245770 during the pre-outburst (blue color), outburst (magenta color) and post-outburst (red color) epochs of our observation campaign are shown. The errors on the flux values are of 1$\sigma$ level. The observation date and corresponding orbital phase ($\phi$) of the neutron star are noted on each profile. The numbers written in the unit of  days with plus and minus signs correspond to days after and before the beginning of the giant X-ray outburst, respectively.  The profiles are plotted with certain offsets for clarity.}
    \label{fig:halpha}
\end{figure}

\begin{figure}
    \vspace{-1.3cm}
    \includegraphics[height=23.7cm,width=8cm]{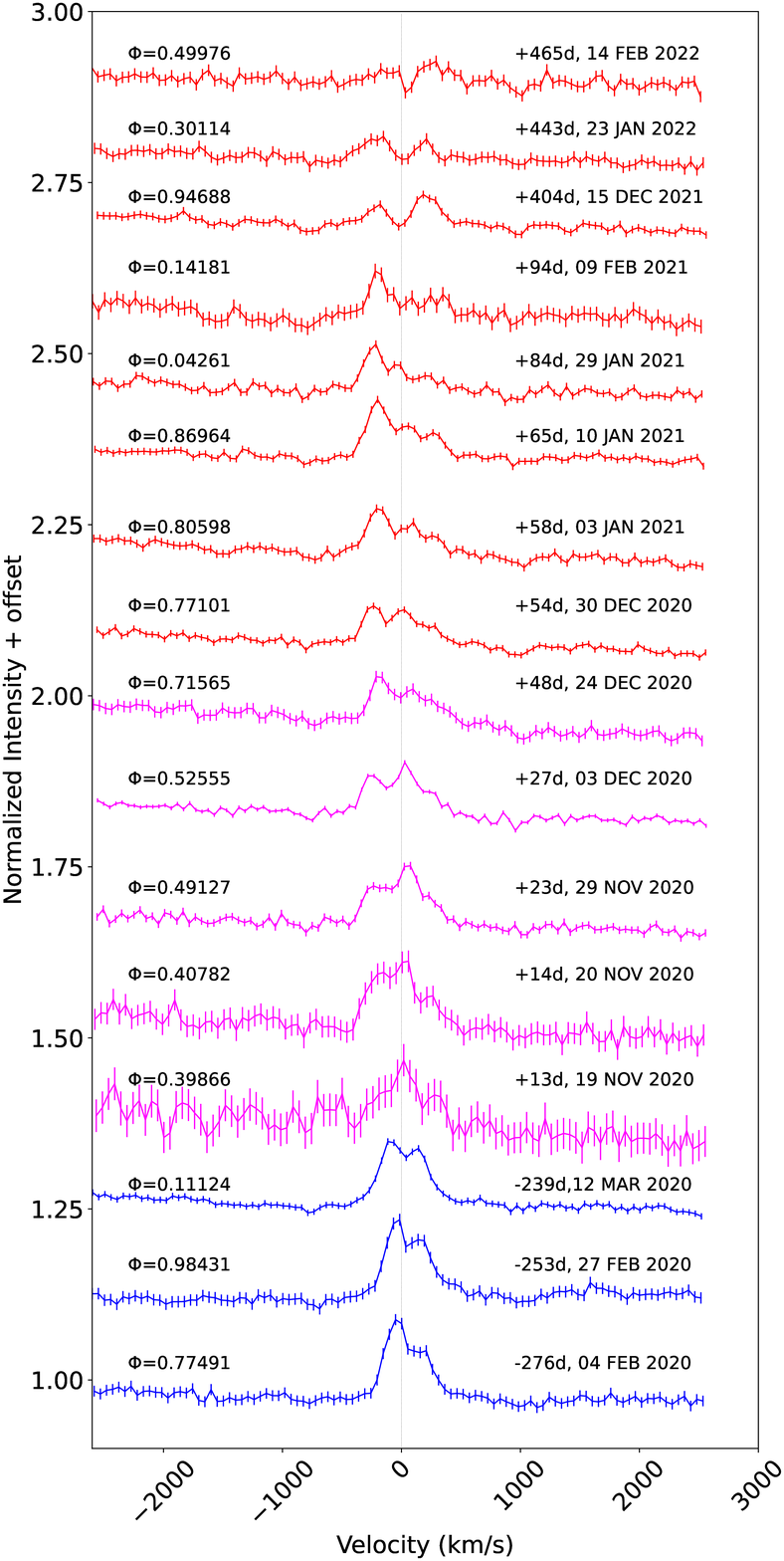}
    \caption{Same as Figure~3, but for He~I (6678~\AA) and with different vertical scale}
      \label{fig:he6678}
\end{figure}

\begin{figure}
    \vspace{-1.3cm}
    \includegraphics[height=23.7cm,width=8cm]{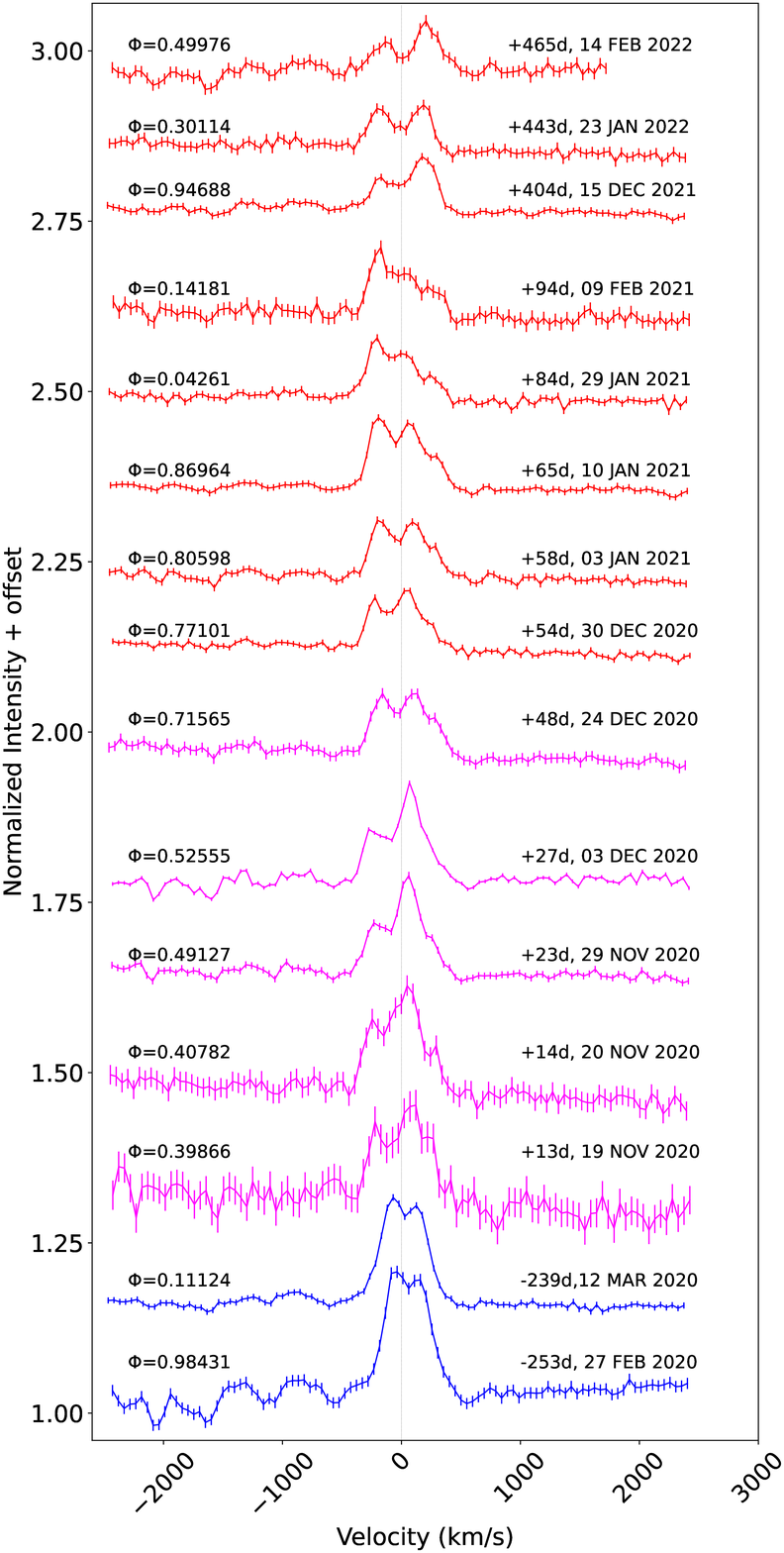}
    \caption{Same as Figure~3, but for He~I (7065~\AA) and with different vertical scale }
    \label{fig:he7065}
\end{figure}

The evolution of the H$\alpha$~(6562.8~\AA), HeI~(6678~\AA), and HeI~(7065~\AA)~ emission line profiles during all our optical observations are shown in Figures~\ref{fig:halpha}, \ref{fig:he6678} \& \ref{fig:he7065}, respectively. In the X-axis, the velocity (in the unit of km/s) is plotted to visualize emission from the approaching and receding parts of the circumstellar disc. The velocities are calculated using the following formula: \newline

Velocity = (Wavelength-$\lambda_{i}$)/$\lambda_{i}$ $\times$3$\times$ 10$^{5}$ , \newline

 where, $i$ = 1, 2, 3 for H$\alpha$, HeI~(6678~\AA), and HeI~(7065~$\AA$), respectively.

The negative velocities exhibit the blue-shifted part, whereas the positive velocity depicts the red-shifted component. In the Y-axis, the normalized intensities with arbitrary vertical offsets are shown. The offsets are given for clarity only. The blue, magenta and red color profiles represent observations during the pre-outburst (quiescent), outburst, and post-outburst quiescent phases, respectively, as in Figure~\ref{fig:spec_evolve}. The date of observation and the corresponding orbital phase of the neutron star ($\phi$) are annotated in each profile. The vertical dotted line corresponds to 0 km/s velocity.

During the pre-outburst X-ray quiescent phase observations, the H$\alpha$ lines are found to be single-peaked and moderately asymmetric with a broad red wing. However, the He~I~(6678~$\AA$) \& He~I~(7065~\AA) line profiles are found to be double-peaked with different intensities. The peak intensity of the blue component is found to be more than that of the red component. During the X-ray outburst, the shape of the H$\alpha$ line profile is found to be opposite to that seen during the pre-outburst phase viz. asymmetric line with a broader blue component. The He~I~(6678~\AA) and He~I~(7065~\AA) lines are found to be complex and multi-peaked during the X-ray outburst.  
  
The H$\alpha$ line profile did not show any significant change during the observations from 2020 December 30 to 2021 February 9 in the post-outburst phase, except for a clear hump-like structure in the blue wing. However, the profile of He~I~(6678~\AA) and He~I~(7065~\AA) lines consisted of  multiple components. During the last three observations (2021 December 15, 2022 January 23 and February 14), the H$\alpha$, He~I~(6678~\AA) and He~I~(7065~\AA) lines are found to be double-peaked with comparable peak intensities.  

\subsection{Investigation of H$\alpha$ line profile}
In the previous subsection, we presented the evolution of the shape of emission line profiles during our observations. In this subsection, we extensively study the H$\alpha$ line parameters.

\subsubsection{Equivalent width}
We calculated the equivalent width ($EW$) of the H$\alpha$ emission lines in the spectra from all epochs of our observations. The $EW$ signifies the strength of the emission/absorption lines. It is a relative and dimensionless measure for the area of emission or absorption line in a wavelength vs intensity plot, and is calculated by the following method:\newline \vspace{0.2cm}
 
 {\Large{ $EW$ = $\sum (1- \frac{F(\lambda)}{F_{c}(\lambda)})\delta \lambda$}} \newline \vspace{0.2cm }
 
where, F($\lambda$) represents the total flux consisting of contributions from the line and the continuum spectrum, and F$_{c}(\lambda)$ represents only the underlying continuum flux at wavelength $\lambda$. The errors in the $EW$s are calculated using the error propagation method which includes continuum flux and wavelength calibration errors. The chosen emission line waveband for H$\alpha$ lines are 6550-6575 \AA. The continuum flux is calculated by taking the median of the flux counts present in the 6517-6550~\AA~ and 6575-6607~\AA~ wavebands. The calculated equivalent widths of H$\alpha$ lines from our observations are quoted in Column~4 of Table~\ref{tab:halpha_par} with 1 $\sigma$ uncertainties. The -ve sign of $EW$ values arises because of the emission line. From this table, it can be seen that the H$\alpha$ equivalent width shows an increasing trend from 2020 February 4 to 2020 March 12 (239 days before the outburst). In absolute terms, the maximum equivalent width of the H$\alpha$ line  that was observed during the 2009 \& 2011 X-ray outbursts was ($\sim$18 \AA; \citet{article_Moritani}). However, we obtained an |$EW$| of $\sim$ 22 \AA \hspace{0.01cm} on 2020 March 12.  

 We also calculated the $EW$ for the He~I~(7065 \AA) line using the same procedure. The chosen emission line waveband for He~I (7065 \AA) is 7052-7078 \AA \hspace{0.01cm} and for the continuum flux, we chose the 7020-7052~\AA~ and 7078-7110~\AA~ wavelength bands. The values are presented in Column~8 of Table~\ref{tab:halpha_par} with 1 $\sigma$ uncertainties. The $EW$ of He~I~(6678~\AA) is not calculated because of low SNR.

\begin{table*}
	\centering
	\caption{Equivalent width, disc radius and other spectral parameters of  H$\alpha$ emission line and equivalent width of the HeI (7065 \AA) emission line.}
	\label{tab:halpha_par}
	\begin{tabular}{lcccccccc} 
		\hline
&  & & & \\
 Date of & MJD ($\phi$) & Days since & $EW$(H$\alpha$)& $\Delta V$ & $FWHM$ & $V/R$  & $EW$ (HeI (7605$\AA$)) & Disc Radius  \\ 
 
 observation &  & outburst &(\AA) & (km/s)& (km/s)  &  & (\AA) & ($R_*$)\\ [4 pt]
\hline 
 & & & & \\
04 FEB 2020 & 58883 (0.78) & $-276 $ &  $-20.67\pm 0.27 $ & $ 183.8\pm 9.6$    & $  369.4\pm 5.0 $    & $ 1.56 \pm 0.01 $  &  & $ 7.62  \pm 0.55 $  \\ [4pt]

27 FEB 2020 & 58906 (0.98) & $-253$ &  $-21.03\pm 0.28 $ & $176.5 \pm 11.4 $    & $353.9 \pm5.0  $     & $ 1.32 \pm 0.01 $    & $- 1.77\pm 0.21$  & $  7.71 \pm 0.69  $  \\ [4pt]

12 MAR 2020 & 58920 (0.11) & $-239$ &  $-22.09\pm 0.17 $ & $ 169.7 \pm11.1 $    & $ 358.9\pm 4.9 $      & $ 0.94 \pm 0.01 $    & $-1.81\pm 0.12 $  & $ 7.96 \pm 0.70   $  \\[4pt]

19 NOV 2020 & 59172 (0.39) & $+13$ & $-17.70\pm 0.77$ & $146.4 \pm 14.1 $    & $ 331.5\pm 6.1 $     & $1.35  \pm 0.15 $    & $-2.11\pm 0.72$  & $ 6.88 \pm 0.96   $  \\ [4pt]

20 NOV 2020 & 59173 (0.40) &  $+14$ & $-17.88 \pm 0.52$ & $147.8 \pm22.3  $    & $323.7 \pm 5.7 $     & $ 0.92 \pm 0.04 $    & $-2.00\pm 0.47 $   & $ 6.93 \pm 1.51  $   \\[4pt]

29 NOV 2020 & 59182 (0.49) & $+23$ &$-15.97\pm  0.19$ & $ 208.0\pm 7.1 $    & $294.6 \pm8.0  $     & $0.24  \pm 0.01 $    & $-1.49\pm 0.15 $   & $ 6.43 \pm 0.33  $   \\ [4pt]

03 DEC 2020 & 59186 (0.52) & $+27$ &$-15.22\pm 0.11$ & $ 196.6\pm 8.5$    & $ 310.8\pm 8.2 $     & $ 0.31 \pm 0.01  $    & $-1.44\pm 0.08$   & $ 6.23 \pm 0.41  $    \\[4pt]

24 DEC 2020 & 59207 (0.71)  & $+48$ & $-13.75\pm 0.27$ & $152.2 \pm 11.0 $    & $ 324.9\pm5.7 $     & $ 0.88 \pm 0.01 $    & $-1.50\pm 0.25 $  & $ 5.82 \pm 0.66  $   \\[4pt]

30 DEC 2020 & 59213 (0.77) & $+54$ & $-12.51\pm 0.15$ & $ 182.9\pm 12.1 $    & $ 321.7\pm 6.7 $      & $ 0.36 \pm 0.02 $   & $- 1.13\pm 0.13$  & $ 5.47 \pm 0.58   $    \\[4pt]

03 JAN 2021 & 59217 (0.80) & $+58$ & $-12.60\pm 0.19$ & $ 185.3 \pm8.7  $    & $ 330.9\pm8.1  $      & $ 0.48 \pm 0.01  $   & $-1.20\pm 0.16$  & $ 5.50 \pm 0.42  $   \\ [4pt]

10 JAN 2021 & 59224 (0.86)  & $+65$ &$-13.69\pm 0.16$ & $195.3 \pm 7.7 $    & $ 337.3\pm6.8 $      & $ 0.33 \pm 0.01 $   & $- 1.33\pm 0.13$  & $5.81  \pm 0.36  $    \\[4pt]

29 JAN 2021 & 59243 (0.04) & $+84$ &$-12.12\pm 0.17$ & $184.4 \pm  9.6$    & $340.4 \pm 6.0$     & $ 0.43 \pm 0.01  $    & $- 0.99\pm 0.15 $  & $ 5.36  \pm 0.46  $   \\[4pt]

09 FEB 2021 & 59254 (0.14) & $+94$ &$-12.24\pm 0.33$ & $ 185.2 \pm 8.3 $    & $336.3 \pm 7.5 $     & $ 0.38 \pm 0.01 $    & $-1.12 \pm 0.28$  & $ 5.39 \pm 0.40  $   \\ [4pt]

15 DEC 2021 & 59563 (0.94) & $+404$ & $- 8.37 \pm 0.17$ & $230.2 \pm 5.1$    & $435.3\pm8.6 $     & $ 1.19 \pm 0.02 $    & $-0.91 \pm 0.15 $  & $  4.19 \pm 0.17   $   \\[4pt]

23 JAN 2022 & 59602 (0.30) & $+443$ &$-8.19 \pm 0.25$ & $ 220.3 \pm 10.5  $    & $ 439.1\pm8.7 $      & $ 1.44 \pm  0.04$   & $- 0.76 \pm 0.21 $  & $ 4.14 \pm 0.37  $   \\ [4pt]

14 FEB 2022 & 59624 (0.49) & $+465$ &$- 7.76 \pm 0.24 $ & $221.0 \pm 9.1  $    & $454.3 \pm 8.8 $      & $ 0.85 \pm 0.02 $   & $-0.66 \pm 0.21$  & $ 3.99 \pm  0.31 $   \\ [4pt]
\hline
 $R_*$ = 1.04 $\times$ 10$^{10}$ m
	\end{tabular}
\end{table*}

 \begin{figure*}
    \hspace*{-2cm}
    \includegraphics[width=22cm, height=20cm]{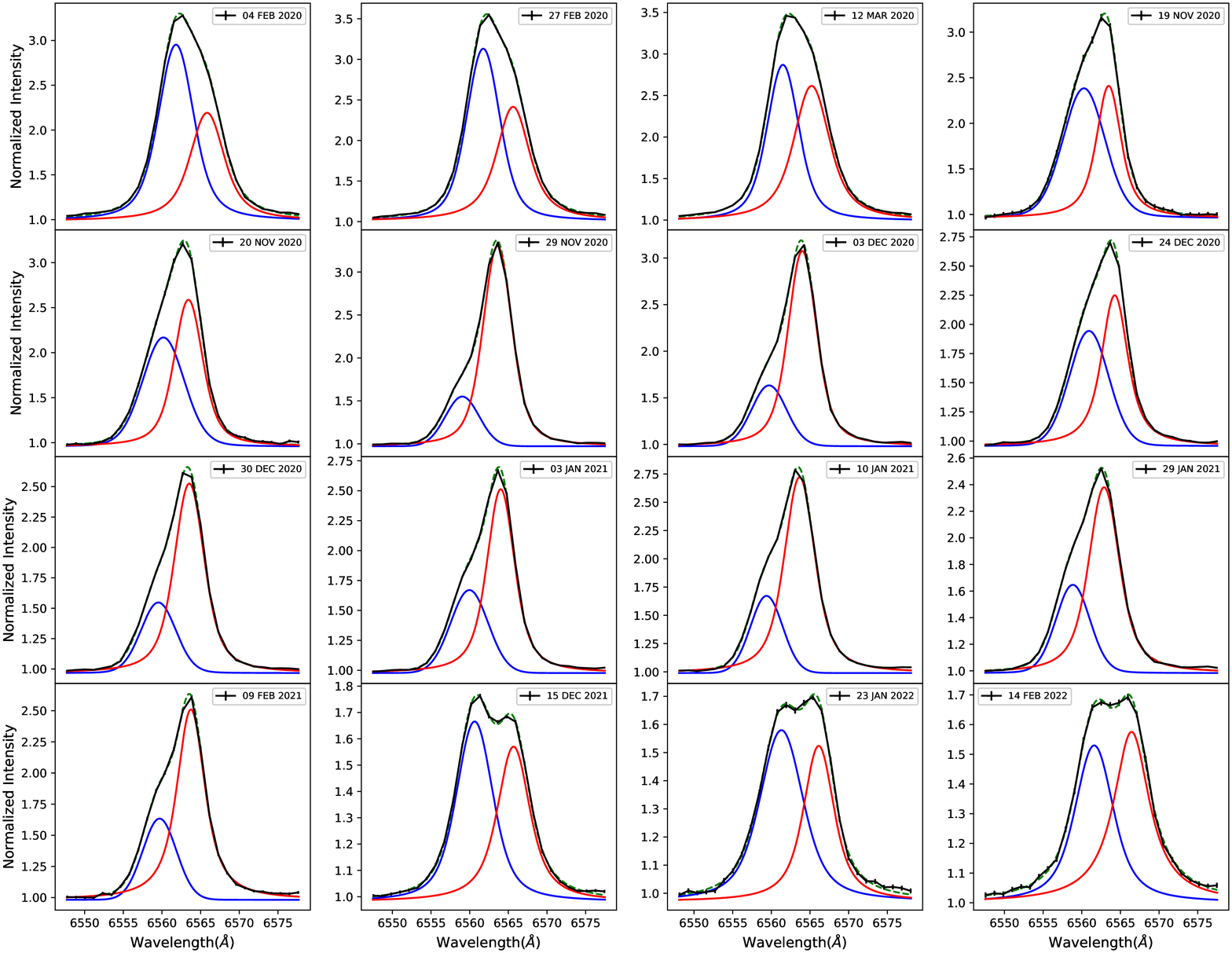}
    \caption{The observed H$\alpha$ line profiles (solid black line) of 1A~0535+262/HD~245770 during our observations and corresponding best-fitted Voigt functions (dashed green line) are shown. The observation dates are quoted on the right of the corresponding profiles. The red and blue color profiles correspond to the red and blue shifted components of the H$\alpha$ line, respectively. }
    \label{fig:voigt}
\end{figure*}
\begin{figure*}
    \includegraphics[width=15cm, height=22cm]{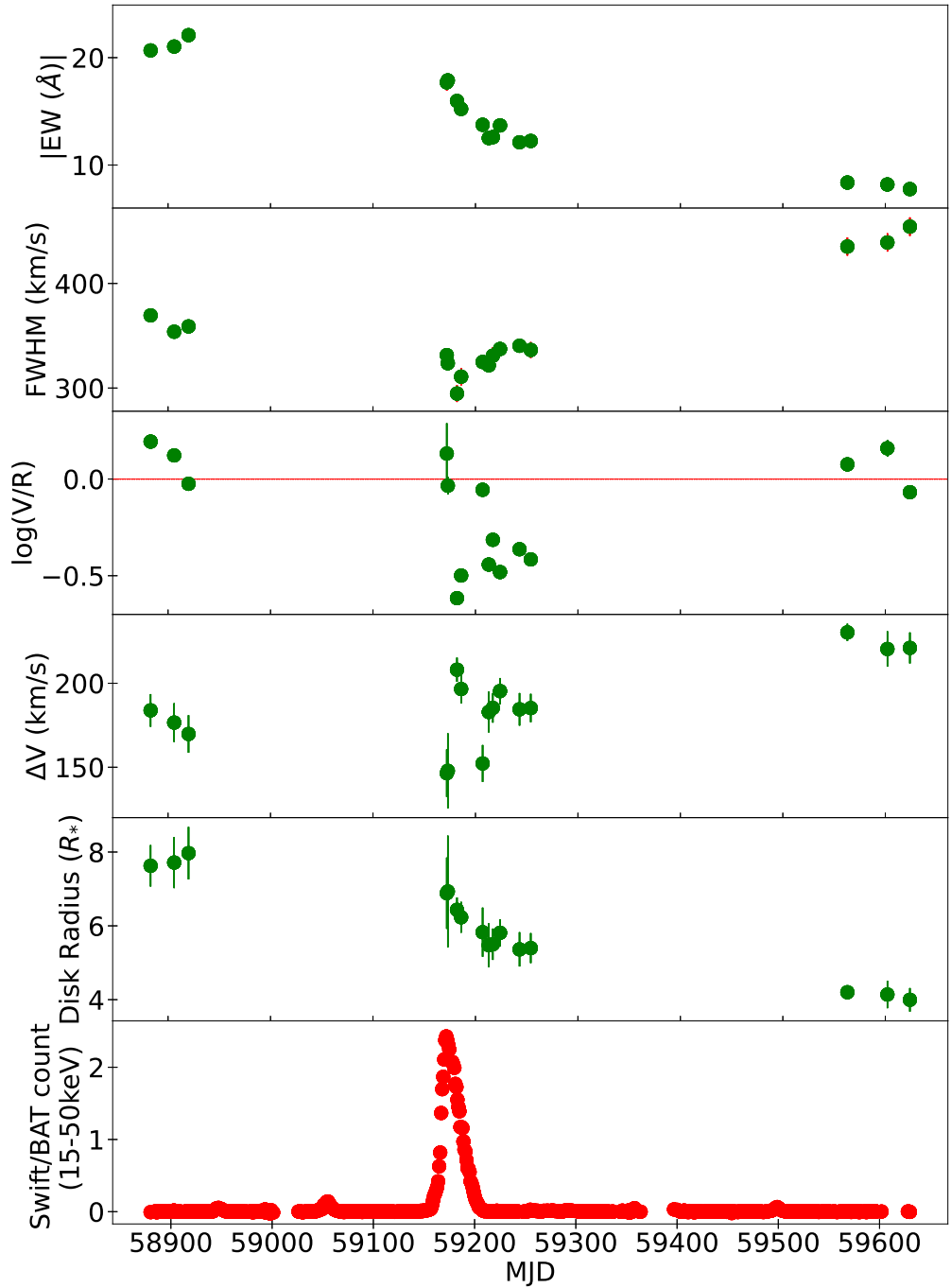}
    \caption{Variation of the H$\alpha$ line parameters of the Be star and the X-ray count rate from the pulsar in 1A~0535+262/HD~245770 system, during our observations are shown. The changes in line equivalent width ($EW$), Full Width at Half Maximum ($FWHM$), $V/R$ ratio, peak separation of fitted components ($\Delta V$), circumstellar disk radius, and Swift/BAT X-ray count rate are shown in panels from top to bottom, respectively. The radius of the Be star is estimated to be ($R_{\ast}$)= 15 $R_{\odot}$ \citep{okazaki2001natural}.}
    \label{fig:variation_halpha}
\end{figure*}

\begin{figure*}
    \includegraphics[width=15cm, height=10cm]{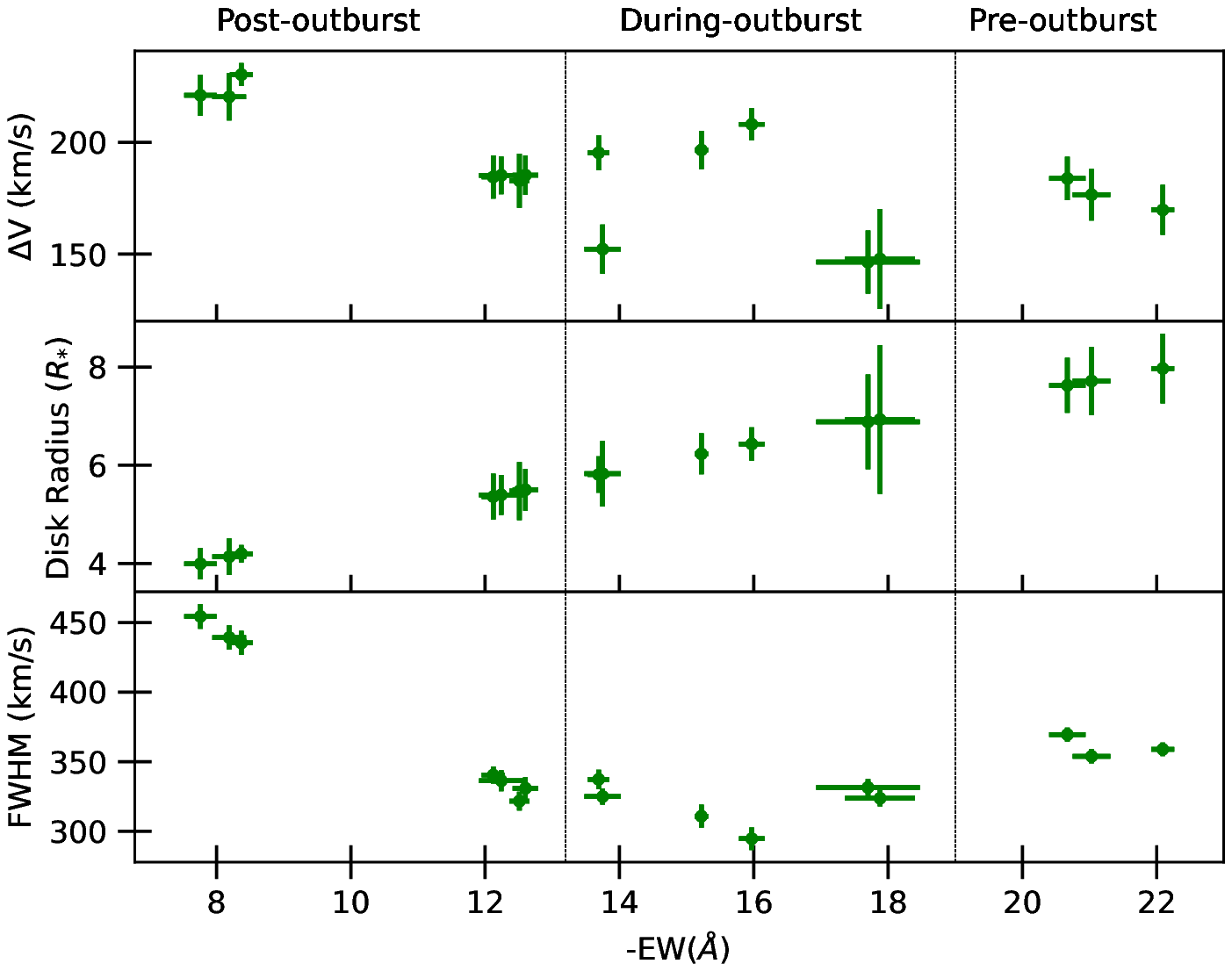}
    \caption{Changes in the separation between the peaks of the blue and red-shifted components of the H$\alpha$ line ($\Delta V$), circumstellar disc radius and the $FWHM$ of the H$\alpha$ line are shown in the top, middle and bottom panels, respectively, as a function of the equivalent width. The regions separated by vertical dotted lines correspond to the observations before, during, and after the giant X-ray outburst, respectively.}
    \label{fig:correlation}
\end{figure*}

\subsubsection{$V/R$ \& $\Delta V$}
Most of the BeXRBs, including 1A~0535+262, show asymmetric H$\alpha$ line profiles \citep{Yan_2011, article_Moritani}. The asymmetries in line profiles are usually mentioned in terms of $V/R$. The $V/R$ referred to the ratio of blue-shifted line flux to the red-shifted line flux. Though most of the observed emission lines in our work are asymmetric and single-peaked, we fitted all the H$\alpha$ emission lines with Voigt profiles to derive the contribution of red and blue-shifted flux. In the beginning, we tried to fit the profiles with Gaussian functions. However, the Gaussian functions did not fit well in the wing regions. The original and fitted line profiles are shown in black and green colors, respectively, in Figure~\ref{fig:voigt}. The red and blue components of the lines are shown with red and blue colors, respectively, in the figure. From this fitting, we extracted several parameters such as central wavelengths, standard deviations, and intensities of blue and red components of the H$\alpha$ line.

The line center of blue and red components are calculated after fitting the Voigt profiles to the H$\alpha$ line. The logarithm of the $V/R$ is plotted in the 3rd panel of Figure~\ref{fig:variation_halpha}. The negative value of $V/R$ implies that the intensity of the red component is more than the blue component and positive value of $V/R$ implies that the blue component is more intense than the red component. It can be noticed from the figure that there is a significant variability occurred in the $V/R$ ratio. On 2020 February 4 (276 days prior to the X-ray outburst), the blue component was $\sim$1.6 times brighter than the red component, whereas on 2020 November 29 (23 days after the onset of the X-ray outburst), the red component was $\sim$5 times brighter than the blue component. After 443 days (on 2022 January 23), again the blue component becomes $\sim$1.4 times brighter than the red component. The $V/R$ variability is an essential factor for tracking the circumstellar disc activity. This variability gives us information on the density distribution of matter in the circumstellar disc. From our observations, we detect a significant variation in $V/R$ that implies an asymmetry in the density distribution in the circumstellar disc. By analysing long-term data from 2005-2009, \citet{10.1111/j.1365-2966.2010.16454.x}, together with \citet{Grundstrom_2007} calculated the periodicity in $V/R$ variability to be approximately 500 days (Figures~5 \& 7 of \citealt{10.1111/j.1365-2966.2010.16454.x}). These results are also consistent with the findings of \citet{Reig2005A&A...440.1079R}. This periodicity, however, is not observed during the 2009 \& 2011 outbursts. Though our data sampling is not suitable enough to check the presence of any periodicity in the $V/R$ variability, significant variability is evident in our $V/R$ measurements.

We can also measure the peak separation ($\Delta V$) which is the difference between the central wavelengths of the red and the blue components of the H$\alpha$ line. In velocity units, it is given by $\Delta V$ = $\Delta  \lambda$/$\lambda_{0}$ $\times$ $c$, where, $c$ is the speed of light, $\Delta  \lambda$ is the wavelength separation and $\lambda_{0}$ is the central wavelength of the H$\alpha$ line. The errors on $\Delta V$ are calculated by using the error propagation method. The peak separation helps in estimating the radius of the Be circumstellar disc \citep{1972ApJ...171..549H}. A small peak separation can be the result of a combined effect of a large disc and a small inclination (face-on) system. Conversely, a large peak separation arises from the combined effect of a small disk and a large inclination (edge-on) system \citep{2017A&A...603A..24M}. The peak separation between the fitted blue and red wings in terms of velocity is plotted in the fourth panel of Figure~\ref{fig:variation_halpha}. 

\subsubsection{Full Width at Half Maximum}

The Full Width at Half Maximum ($FWHM$) is the width of the line at half of the peak flux levels of the emission or absorption line. The $FWHM$ can be expressed in the velocity unit to estimate the radial velocity of the H$\alpha$ emitting particles in the disc with respect to the observer. In this work, the $FWHM$ is calculated by fitting a single Gaussian function over the entire H$\alpha$ emission line using a python routine that fits the overall line shape well. The variation of the $FWHM$ with time is shown in the second panel of Figure~\ref{fig:variation_halpha}. The $FWHM$ is found to vary in the range of 6.44-9.93 \AA \hspace{0.01cm} (294-454 km/s) during our observations.

\subsubsection{ Estimation of H$\alpha$  emitting region}

We can calculate the size of the disc using the H$\alpha$ line parameters. \citet{1972ApJ...171..549H} demonstrated that the size of the H$\alpha$ emitting region of the circumstellar disc can be estimated by using the peak separation ($\Delta V$) of the double-peaked H$\alpha$ emission line, assuming Keplerian velocity distribution of matter in the disc. \citet{1989Ap&SS.161...61H} also demonstrated that for single-peaked profiles, there exists a linear relation between $\Delta V$, $ V\sin i$ and equivalent width (EW) which is given by : 

\begin{equation}
log(\frac{\Delta V}{2V \sin i})= -a \times \log (-EW/ \text{\AA}) + b
\end{equation}

where, $a$ and $b$ are the slope and intercept of the plot log($\frac{\Delta V}{2V \sin i}$) vs log(-$EW$/\AA), and $i$ is the inclination angle of the disc with respect to the observer. For a sample of BeXRBs, \citet{10.1093/mnras/stw2354} calculated the values of $a$ and $b$ to be 0.334 and 0.033, respectively. Using the values of $\Delta V$ from our H$\alpha$ line profile fitting, we calculated $V\sin i$ by using the following equation :\newline
\begin{equation}
V~\sin i= \frac{\Delta V}{2 \times 10^{-0.334\times \log(-EW/\text{\AA} ) + 0.033}}  
\end{equation}
 
  The average value of $V\sin i$ is estimated to be 236.84$\pm$13.55 km s$^{-1}$ which is in agreement with the reported value of 225 $\pm$10 km s$^{-1}$ (\citealt{reig2011x} \& references therein). The velocity separation between the red and blue shifted components of emission lines can be used to calculate the radius of the emitting region \citep{1972ApJ...171..549H}. The relationship between the peak separation and the rotational velocity of the gas particles at the emitting region is  $\Delta V$= 2$V_{rot}\sin i$. Then the size of H$\alpha$ emitting region can be derived by :\newline \vspace{0.2cm} 
\begin{equation}
R_{d}= (2V \sin i / \Delta V)^{j} \epsilon R_{\ast} 
\end{equation}

where, $j$=2 for Keplerian rotation, $R_{\ast}$ is the Be star radius and $\epsilon$ is a dimensionless parameter that considers several effects that would over-estimate the disc radius ($\epsilon$=0.9$\pm$0.1 ; \citealt{Zamanov2013A&A...559A..87Z}). For 1A~0535+262, the radius of the Be companion star is 15 $R_{\odot}$ \citep{okazaki2001natural}. We estimated the radius of the circumstellar disc during our observations and plotted this in the fifth panel of Figure~~\ref{fig:variation_halpha}. From the figure, it can be noticed that before the giant X-ray outburst, the size of the disc was increasing until the onset of the outburst and decreased continuously after the onset to a value of 4$R_{*}$.

\subsection{Correlations between $EW$s and other parameters of H$\alpha$ line}

Figure~\ref{fig:correlation} depicts the variation in the H$\alpha$ line parameters such as the peak separation (top panel), disc radius (middle panel), and $FWHM$ (bottom panel) against the line equivalent width ($EW$). Two vertical dashed lines separate the data points into three sections. The section, in the beginning, corresponds to the post-outburst epochs, whereas the sections in the middle and extreme right correspond to the outburst and pre-outburst epochs. An anti-correlation between the peak separation ($\Delta V$) and the equivalent width of the H$\alpha$ line can be seen in the top panel of the figure throughout the pre- and post 2020 giant X-ray outburst. These parameters do not follow any specific trend during the outburst. A clear correlation between the disc radius with the $EW$ can be seen during all the epochs of observations.  An anti-correlation is seen between the $FWHM$ and  the $EW$ in the post-outburst phase, though no clear dependency is found  before and during the outburst.

\subsection{Long-term optical and X-ray variation of the system}
\begin{figure*}
    \hspace*{0cm}
    \includegraphics[width=6.3in, height=3.9in]{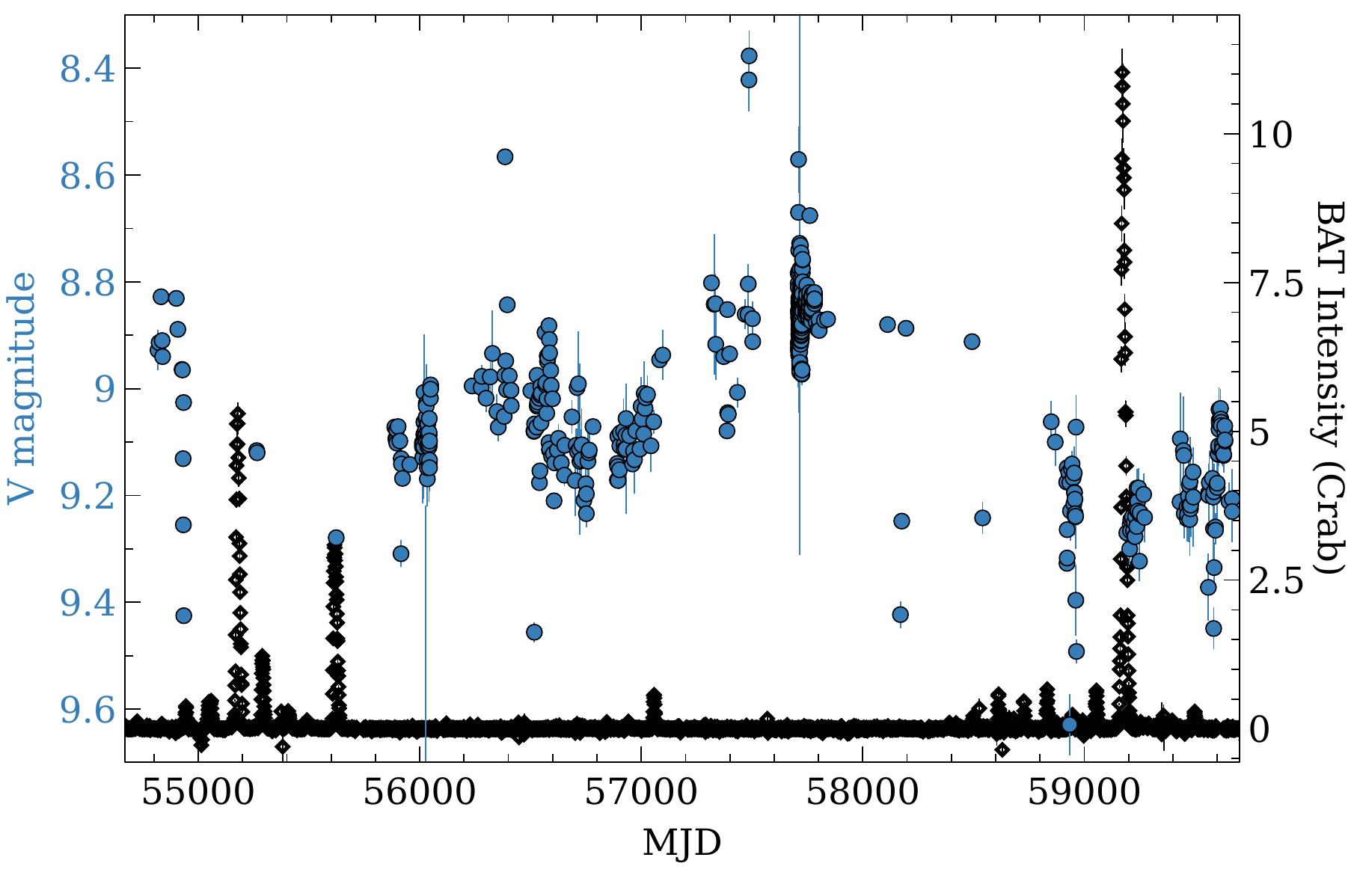}
    \caption{Long-term time evolution of $V$-band magnitude and X-ray flux of 1A~0535+262/HD~245770 obtained from AAVSO international database and Swift/BAT (in 15-50 keV energy range), respectively.  The cyan-filled circles represent $V$-band magnitudes while the open black diamonds  represent BAT intensity in Crab units.} 
    \label{fig:aavso}
\end{figure*}

We also studied the long term evolution of the Be star in the $V$-band using the publicly available data from the American Association for Variable Observers (AAVSO) international database together with the X-ray monitoring light curve of the pulsar from Swift/BAT in 15-50 keV range (Figure~\ref{fig:aavso}). It is done to understand the companion properties before, during, and after the X-ray outbursts observed between MJD 54670 (2008 July 23) and 59700 (2022 May 1). Figure~\ref{fig:aavso} shows a dimming of the star before the 2020 giant outburst where the $V$-magnitude was highest at $>$9.4 around MJD 58965. At the same time, we also see an increase in the equivalent width of the H$\alpha$ line (first three points in Figure~\ref{fig:variation_halpha}, corresponding to the pre-outburst phase). A similar kind of behavior is also seen before the 2009 giant outburst \citep{Camero_Arranz_2012, Yan_2011}.

\section{X-ray results}

Following the onset of the giant X-ray outburst, AstroSat observed 1A~0535+262 during the rising phase of the outburst on 2020 November 11 (MJD 59165), as shown in  Figure~\ref{fig:swiftlc}. The pulsar was observed for an exposure of $\sim$94~ks in a broad energy range using the SXT, LAXPC, and CZTI instruments onboard the satellite. We carried out timing and spectral analysis of data from all the detectors and presented it in this work.

\begin{figure}
    \centering
    \includegraphics[width=10cm, height=8cm, angle=-90 ]{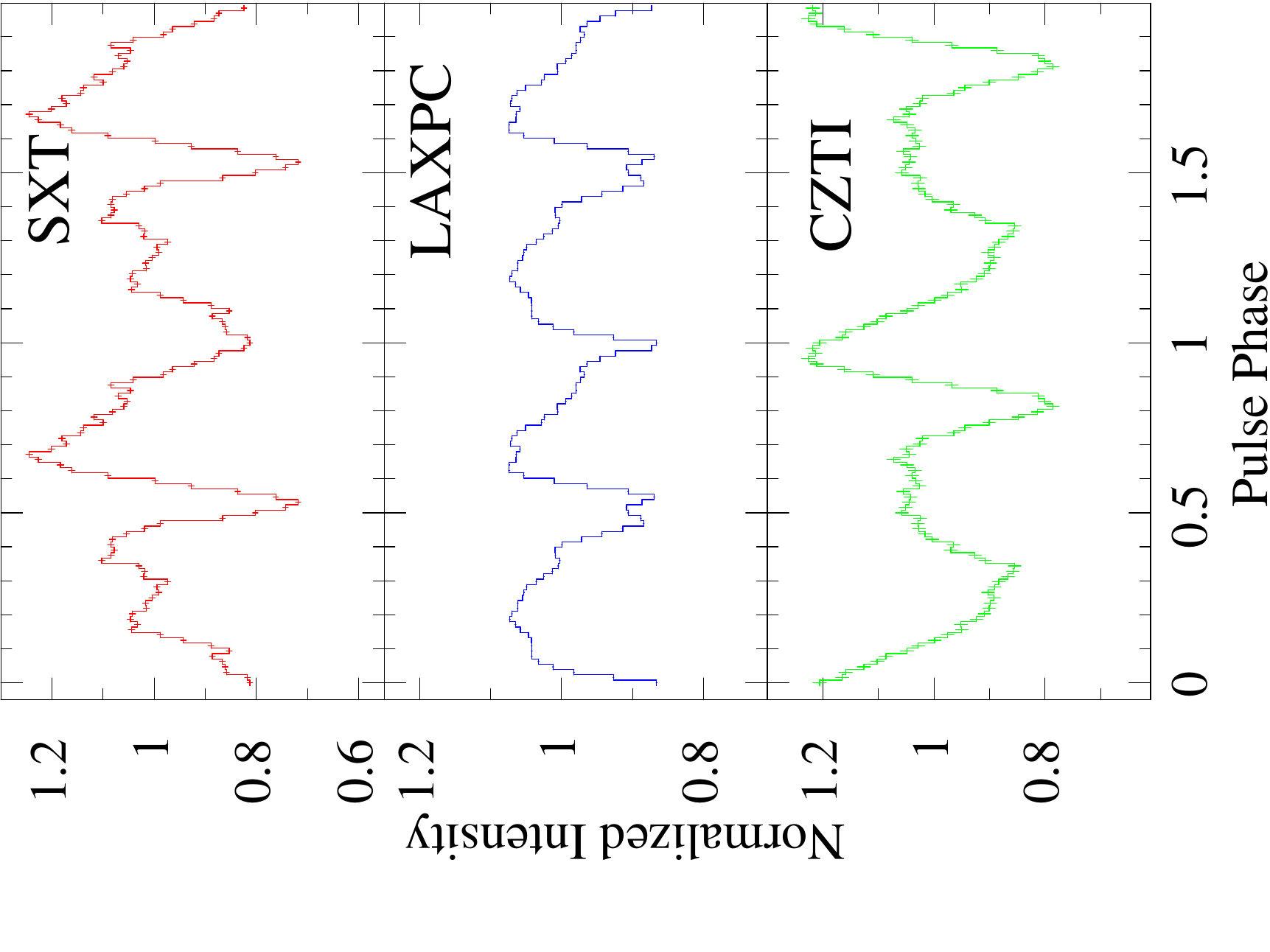}
    \caption{Average pulse profiles of 1A~0535+262 obtained from barycenter corrected and background subtracted light curves of AstroSat's SXT (0.3-7 keV; Red), LAXPC (3-80 keV; Blue) and CZTI (20-110 keV; Green) from top to bottom panels, respectively. The errors on the data points are of 1$\sigma$ level. Two cycles are shown for clarity.}
    \label{fig:pp_broad}
\end{figure}

\begin{figure*}
    \centering
    \includegraphics[width=13cm, height=18cm,angle=-90]{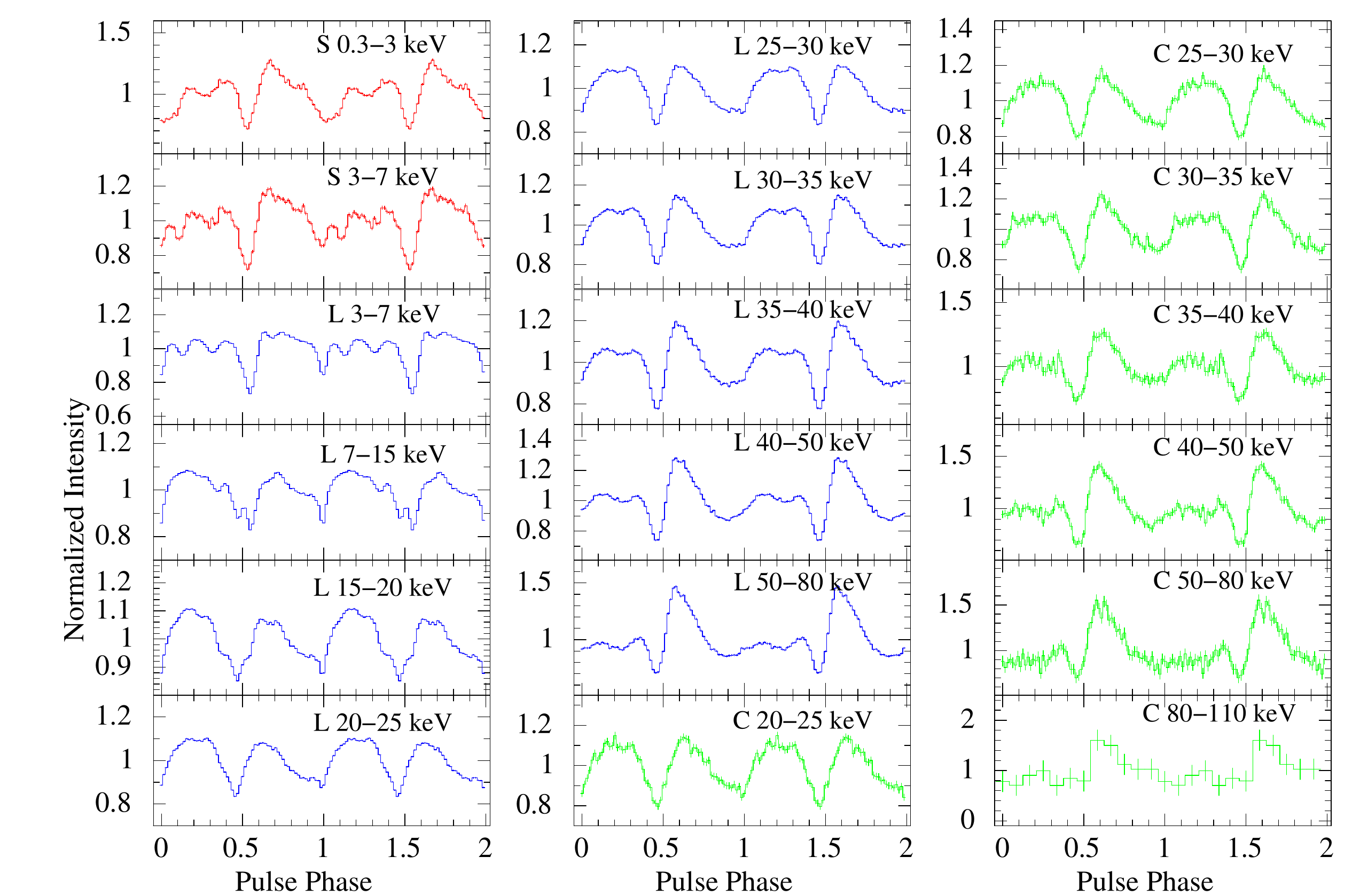}
    \caption{Energy-resolved  pulse profiles of 1A~0535+262 obtained from barycenter corrected and background subtracted light curves of AstroSat's SXT (S; Red), LAXPC (L; Blue), CZTI (C; Green). The errors on the data points are of 1$\sigma$ level. Two cycles are shown for clarity.}
    \label{fig:erpp}
\end{figure*}

\subsection{Timing analysis}
The source and background light curves for SXT instrument in the 0.3-7 keV energy range were extracted with a bin size of 2.4~s from the reprocessed and merged SXT event file. For the LAXPC and CZTI instruments, the source and background light curves were extracted with a bin size of 0.1~s in 3–80 keV and 20–110 keV ranges, respectively. Barycentric corrections were applied to the background subtracted light curves using the \textit{as1bary}\footnote{http://astrosat-ssc.iucaa.in/?q=data\_and\_analysis} tool. The $\chi^{2}$ maximization technique \citep{1987A&A...180..275L} was applied to search for pulsations in the LAXPC light curve by using the \textit{efsearch} task of FTOOLS package. The pulse period of the pulsar was estimated to be $\sim$103.548(4)~s. Using this period, the background subtracted and barycenter corrected SXT, LAXPC, and CZTI light curves were folded in the energy ranges of 0.3-7 keV, 3-80 keV, and 20-110 keV, respectively. While generating pulse profiles using the {\it efold} task of  FTOOLS, a suitable epoch was used to align the minimum of the profile to phase zero. Corresponding folded pulse profiles are shown in the top, middle, and bottom panels of Figure~\ref{fig:pp_broad}, respectively. The pulse profiles are found to be double-peaked in the SXT, LAXPC, and CZTI energy ranges. The shape of the pulse profile, however, differs in soft and hard X-ray ranges. A phase shift of $\sim$0.1 is seen in the CZTI (hard X-ray) profile compared to the SXT and LAXPC profiles.

To further investigate the energy dependence of the pulse profile, we extracted light curves in narrow energy bands from the SXT, LAXPC, and CZTI data. Light curves in 0.3–3 keV and 3–7 keV ranges were extracted for SXT data, whereas light curves in seven energy ranges (20–25 keV, 25-30 keV, 30-35 keV, 35-40 keV, 40-50 keV, 50-80 keV, and 80-110 keV) and nine energy ranges (3-7 keV, 7-15 keV, 15-20 keV, 20-25 keV, 25-30 keV, 30-35 keV, 35-40 keV, 40-50 keV, and 50-80 keV) were extracted from the CZTI and LAXPC data. Pulse profiles were generated from the above energy resolved SXT, LAXPC, and CZTI light curves by using an estimated spin period of the neutron star and are shown in Figure~\ref{fig:erpp}. The same epoch, as used earlier, was used to generate the energy resolved profiles. From the figure, it can be seen that the pulsation in the light curve can be seen up to 110 keV. Apart from the primary dip in the 0.45--0.55 phase range, several small and narrow dip like features (secondary dips) were observed in the pulse profiles of the pulsar in 0.3-3 keV and 3-7 keV ranges (Figure~\ref{fig:erpp}). All of the secondary dips, except the one in the 0.9--0.0 pulse phase range, further diminished and subsequently were found to be absent in the profiles in high energy ranges. The secondary dip in the 0.9--0.0 phase range, however, became prominent making the profiles double-peaked up to $\sim$35 keV. Apart from the change from single-peaked to double-peaked, there appears to be a marginal shift in the phase of the profiles beyond 15 keV. The shape of the profile keeps on changing with an increase in energy. Beyond 35 keV, the second peak gets diminished in the profiles from the LAXPC and CZTI data, eventually becoming consistent with a single-peaked at higher energies. The observed change in the shape of the pulse profiles signifies its evolution with energy.  Apart from the appearance and disappearance of the secondary dips causing changes in the pulse profiles in various energy ranges, it can be noticed that there is an indication of the appearance of a minor dip in the 0.2--0.3 pulse phase range in the pulse profiles beyond 25 keV. The appearance of this dip coincides with the reduction of the intensity of the first peak.

\begin{figure}
    \hspace*{-1cm}
    \includegraphics[height=5cm, width=10cm]{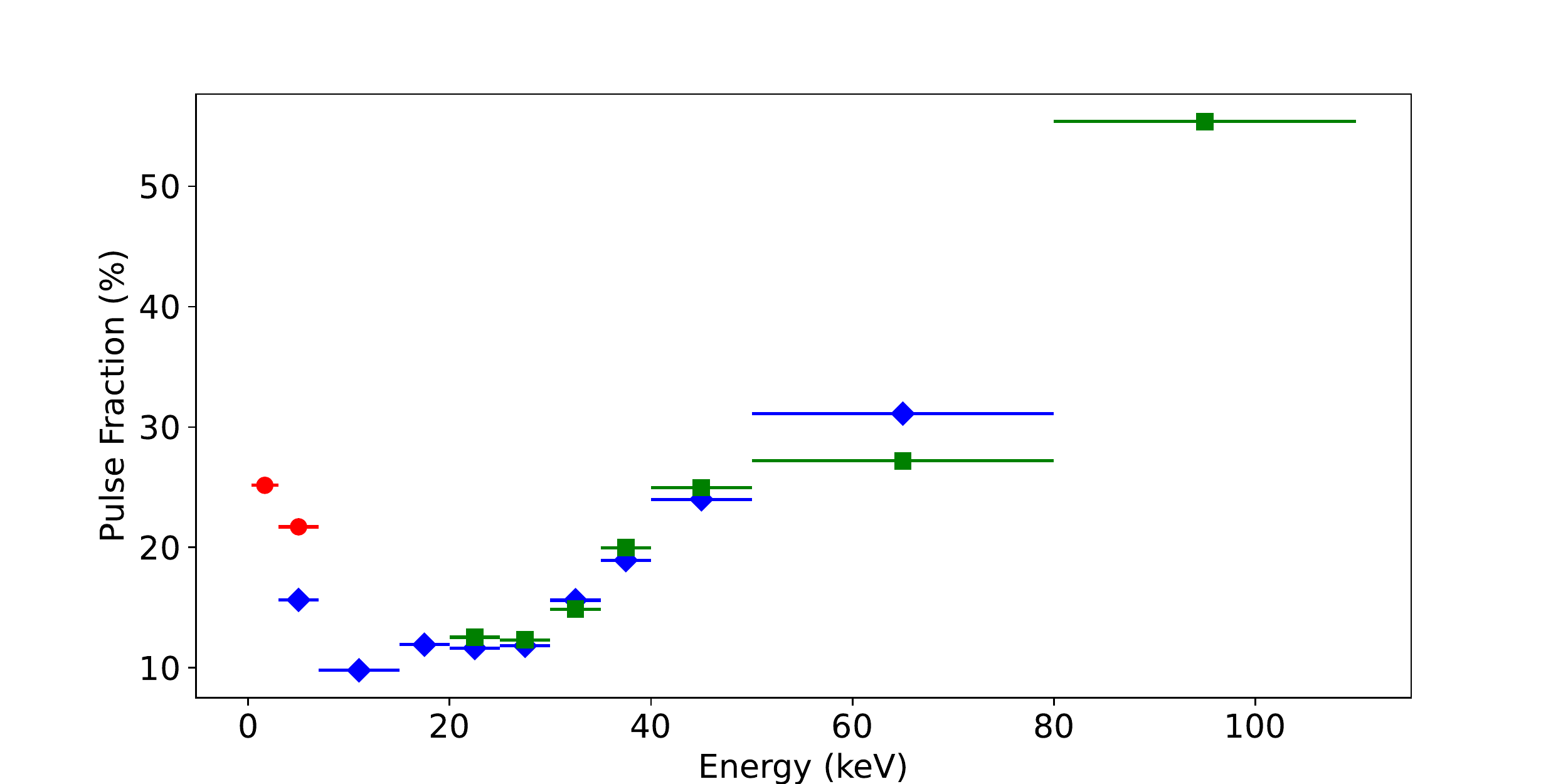}
    \caption{Energy-resolved pulse fraction of 1A~0535+262 for SXT (red), LAXPC (blue), and CZTI (green). Energy bin sizes are depicted by horizontal bar.}
    \label{fig:pf}
\end{figure}

To further investigate the energy dependency of the pulse profiles, we calculated a quantity called the pulse fraction (PF). The PF is defined as the ratio between the difference and the sum of maximum (I$_{max}$) and minimum (I$_{min}$) intensities in the pulse profile. PF =  (I$_{max}$ - I$_{min}$)/ (I$_{max}$ +I$_{min}$). In Figure~\ref{fig:pf}, the variation of pulse fraction with energy is shown. We found that the pulse fraction decreases from $\sim$25\% to 9\%  from 0.3 to $\sim$ 15 keV. Beyond $\sim$15 keV, the pulse fraction of the pulsar showed an increasing trend with energy. In the energy range of 15-110 keV, the pulse fraction increases from 10\% to 55\%.

\subsection{Spectral analysis}
The spectral analysis of the AstroSat observation of 1A~0535+262 was performed using XSPEC v-12.9.0 \citep{1996ASPC..101...17A} package. The SXT spectrum was obtained in the 0.7-7.0 keV energy band and was grouped to have a minimum of 30 counts per bin by using the {\it grppha} task of  FTOOLS. The extracted LAXPC spectrum was grouped to have a minimum of 30 counts per bin. We found that the LAXPC data were background dominated above 60.0 keV. Therefore,  the 3.0-60.0 keV range was considered in the spectral fitting. For the same reason, 30.0-90.0 keV range data were chosen for the CZTI instrument. The CZTI spectrum was grouped to have a minimum of  30 counts per bin.  We carried out spectral fitting in the 0.7-90.0 keV range using simultaneous data from the SXT, LAXPC, and CZTI instruments. To include the effect of Galactic absorption, we used {\it tbabs} model component with abundances from \citet{2000ApJ...542..914W} and cross-sections as given by \citet{1996ApJ...465..487V}.

\begin{figure}
    \hspace*{-1cm}
    \includegraphics[]{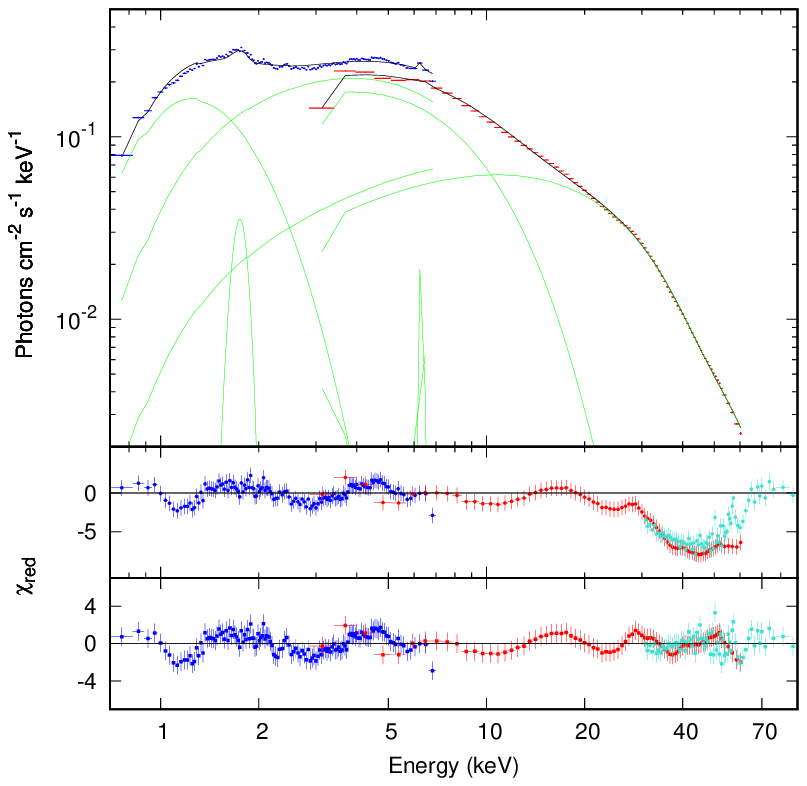}
    \caption{Top panel : Broadband AstroSat (SXT+LAXPC+CZTI) spectra of 1A~0535+262 are shown along with the best-fit spectral model. The red, blue, and cyan  color data points correspond to the SXT, LAXPC, and CZTI instruments, respectively. The cut-off power-law, two blackbody components, and Gaussian lines are shown in green colors. The best-fit spectrum in the 0.7-90.0 keV range is shown in black color. The middle and bottom panels show the residuals without and with the CRSF component in the spectral model, respectively.}
    \label{fig:spec}
\end{figure}

While performing simultaneous spectral fitting, a relative normalization constant was taken into account in order to address the calibration differences between the three instruments. This factor was fixed to 1 for the SXT  data and allowed to vary for the LAXPC and CZTI data. In the beginning, a power law with an exponential cutoff continuum model was used to fit the spectrum. The presence of positive residuals in the soft X-ray range allowed us to add a blackbody component to the continuum model. While fitting the data with the above model, however, another blackbody component was required to fit the spectrum well in the soft X-ray ranges. Two blackbody components: a hotter one (kT $>$ 1 keV) and a cooler one (kT $<$ 1 keV) were added for the soft X-ray excess observed in the low energy range of the spectrum, as reported in the previous studies \citep{Jaisawal2020ATel14179....1J, 2021ApJ...917L..38K}. We used \texttt{gain fit} command to the SXT data for gain correction with a fixed slope of 1 to flatten the residuals at 1.8 and 2.2 keV. Despite the gain correction, we also see a feature at $\sim$1.8 keV which is fitted with a Gaussian emission line. As the feature is related to the detector calibration, it is not considered intrinsic to the source. Gain correction is also applied to the LAXPC data as suggested in \citet{2021JApA...42...32A}. A weak iron emission line was detected in the spectrum and modeled with a Gaussian function at $\sim$6.4 keV. While fitting the data with the above model, negative residuals were found in the 40.0-50.0 keV range. An addition of a Gaussian absorption (\textit{GABS}) model with the central energy of $\approx$46.3 keV to the above model improved the spectral fitting significantly.  While fitting the line width was fixed at 10 keV, as was done in \cite{2021ApJ...917L..38K}. This absorption feature is considered as the cyclotron resonance scattering feature in the spectrum of the pulsar.  The broad-band spectrum of the source along with the best-fit spectral model is shown in the top panel of Figure~\ref{fig:spec}. The residuals obtained without and with the addition of the Gaussian absorption component at 46.30 keV in the spectral model are shown in the middle and bottom panels of the figure, respectively. The spectral parameters obtained from our fitting are quoted in Table~\ref{tab:specfit}.

\begin{table}
\caption{Best-fit parameters obtained from spectral analysis of AstroSat observation of 1A~0535+262}
 \label{tab:specfit}
\centering
\begin{tabular}{ lcc } 
\hline
\hline
Model & Parameters &  Parameters Values \\ [4 pt]
\hline 
 TBabs   &    $\rm N_{H} (10^{22}~cm^{-2}$) & $0.64_{-0.03}^{-0.03}$ \\[6 pt]
Cutoffpl & Photon Index ($\Gamma$)  &  $-1.08_{-0.14}^{+0.12}$        \\ [6 pt]
         & $\rm E_{cut}$ (keV)           &  $10.02_{-0.34}^{+0.33}$           \\ [6 pt]
         & norm ($10^{-2}$             & $1.62_{-0.4}^{+0.5}$        \\[6 pt]
Bbodyrad1 & kT (keV)  & $0.28 _{-0.01}^{+0.01}$                   \\[6 pt]
          & norm      & $ 5758 _{-567}^{+626}$                  \\[6 pt]
Bbodyrad2 & kT (keV)  & $2.25 _{-0.03}^{+0.03}$                   \\[6 pt]
          & norm     & $ 49.67 _{-2.2}^{+2.0}$                  \\[6 pt]
Gabs      &  $E_{\mathrm{cyl}}$ (keV)    & $  46.30_{-0.74}^{+0.81}$       \\[6 pt]
          &  width $\sigma$ (keV)      &  10.00 (fixed)      \\[6 pt]
          &  depth $\tau$          & $ 7.58_{-0.73}^{+0.76}$       \\[6 pt]
Gaussian & E (keV)  &   $6.34_{-0.03}^{+0.03}$  \\ [6 pt]
          & width $\sigma$ (keV) & $0.06_{-0.03}^{+0.04}$ \\ [6 pt]
          & norm ($10^{-2}$) & $0.54_{-0.1}^{+0.1}$ \\ [6 pt]
Unabsorbed flux & ($\rm 10^{-8}~erg~cm^{-2}~s^{-1}$) & $ 7.25 \pm 0.95$  \\[4 pt] 
Fit Statistics  &    $\rm \chi_{red}^{2}/d.o.f.$ &        1.1/781            \\
\hline
\end{tabular}
\end{table}

\section{Discussion}

We have carried out optical and X-ray studies of 1A~0535+262 across its 2020 giant X-ray outburst using Mnt. Abu Infrared Observatory and AstroSat observations, respectively. The neutron star in the BeXRB systems is known to show X-ray outbursts (Type~I) at the periastron passage, although the X-ray activity depends on the evolution of the circumstellar disc of the companion Be star at the truncation radius \citep{okazaki2001natural}. At the periastron passage, if the circumstellar disc is sufficiently evolved, the neutron star accretes matter from the disc and shows an increase in X-ray luminosity, lasting for about 20\% of the binary orbital period. The loss of matter from the circumstellar disc also affects the Be star's overall optical and infrared emission. During giant X-ray outbursts, on the other hand, the X-ray luminosity gets enhanced by several orders of magnitude and lasts for a few orbital periods, unlike Type~I outbursts. It becomes crucial to probe the dynamics of the Be star and its circumstellar disc around strong and quiescent X-ray states. We have studied the evolution of H$\alpha$ and He emission lines in the 1A~0535+262/HD~245770 system before, during, and after the recent giant X-ray outburst in 2020 covering a period of two years to understand the changes in the circumstellar disc as discussed in this section.

\subsection{Evolution of H$\alpha$ line profile}
In the present work, we investigated the evolution of the H$\alpha$ emission line profile before, during, and after the 2020 giant X-ray outburst. The evolution of the H$\alpha$ emission line profiles is shown in Figure~\ref{fig:halpha}.  Before the X-ray outburst, the  H$\alpha$ line profile was single-peaked and moderately asymmetric with a broad red wing. During the X-ray outburst, the line profiles were asymmetric and single-peaked but exhibited the opposite nature to that of the pre-outburst phase, i.e., the blue wing is broader than the red wing. After the X-ray outburst, the line profile did not change much for up to one month except for a hump-like structure in the blue wing, and then within one year, the profile evolved into a double-peaked shape.

The evolution of the H$\alpha$ emission line was investigated during the 2009 and 2011 giant X-ray outbursts of 1A~0535+262/HD~245770 system \citep{Camero_Arranz_2012}. The observations were carried out at a spectral resolution of R$\sim$5000. The emission lines were observed as single-peaked, though marginally asymmetric before and during the outburst. During the 2009 outburst, the H$\alpha$ line was found to be asymmetric with a broader blue wing, whereas during the 2011 outburst, the profile had a broader red wing (left and the right panels of Figure~4 of \citealt{Camero_Arranz_2012}). On comparing the observed H$\alpha$ line profile of the Be star during and before the 2020 giant X-ray outburst, it can be seen that it resembles the profile of the H$\alpha$ line during the 2009 outburst. Our observations indicate that the emission line profiles after the 2020 giant outburst are red-shifted with a broader blue wing. However, before the outburst, the profiles of the H$\alpha$ line were blue shifted with a broader red wing. The studies of H$\alpha$ emission line with higher spectral resolution ($R\sim$ 30000-60000), around the 2009 X-ray outburst, showed that the basic structure of the emission line (Figure~2 of \citealt{article_Moritani}) is consistent with that in this work as well as in \citet{Camero_Arranz_2012} though some absorption features were also reported in the latter paper. The absorption features were detected in the spectra due to the higher resolution of the instrument. These features indicate the complex structure of the Be circumstellar disc. \citet{Camero_Arranz_2012} suggested the warping of the Be disc as the reason behind the observed H$\alpha$ line profile structure. From a long term observation of 4U 0115+634/V635~Cas, \citet{2001A&A...369..108N} and \citet{2007A&A...462.1081R} suggested that a precessing warped Be circumstellar disc caused the change in the profile from a normal double-peaked to a single-peaked profile during the giant X-ray outburst. Similar behaviour was also observed by \citet{10.1093/pasj/63.4.L25} during the 2009 giant outburst in 1A 0535+262.

During the 2020 giant outburst, we estimated the size of the Be circumstellar disc using the observed H$\alpha$ emission line parameters. The evolution of the size of the Be disc during our observations is shown in the fifth panel of Figure~\ref{fig:variation_halpha}. It is clear that the disc size was largest during the observations prior to the 2020 giant X-ray outburst. The size of the disc started to decrease since the beginning of the outburst and remained small during the observations even after the outburst. The disc radius before the giant outburst on MJD 58920 (2020 March 12) was $\sim 8.27 \times 10^{10}$ m which is equivalent to $\sim$7.62 times the Be star radius. Using the viscous decretion model \citep{2001A&A...369..108N}, the calculated 4:1 resonance truncation radius for 1A~0535+262 is $7.27\times 10^{10}$ m. The Roche lobe radius at periastron passage is estimated to be $7.08\times10^{10}$m \citep{okazaki2001natural}]. The viscous decretion model suggests that the truncation radius limits the size of the Be circumstellar disc as the viscous and tidal torque exactly balance each other. Later, \citet{okazaki2007interaction} extended the viscous decretion model for highly eccentric and misaligned (spin axis of the Be star and the normal to the binary plane are not parallel) systems and showed that the truncation is less efficient when the inclination (i) is $\geq$ 60$^{\circ}$. As the estimated Be disc size in 1A~0535+262 (prior to the giant X-ray outburst) is larger than the truncation radius, the extended viscous decretion model confirms that the Be circumstellar disc and the binary plane of the BeXRB are highly misaligned. As suggested by \citet{10.1111/j.1365-2966.2011.19231.x} and \citet{10.1093/pasj/65.2.41}, this highly misaligned disc could be warped before the giant X-ray outburst in 2020 for the neutron star to capture a huge amount of matter. The warping of the disc before the 2009 giant X-ray outburst in 1A~0535+262 has also been considered \citep{10.1093/pasj/63.4.L25}.

As shown in Figure~\ref{fig:variation_halpha}, the H$\alpha$ line equivalent width (size of H$\alpha$ emitting region) was maximum before the giant X-ray outburst. However, during and after the outburst, the equivalent width decreased from its pre-outburst values. A decrease in the value of equivalent width during and after the outburst suggests that a significant amount of matter is lost from the circumstellar disc during the outburst. By observing the change in the quantized IR excess flux, \citet{10.1111/j.1365-2966.2004.07743.x} suggested that in 1A~0535+262, the Be disc grows and decays in a quasi-cyclic period of 1500 d. Considering this, the disc size is anticipated to grow or remain comparable (in case no mass is ejected from the Be star). However, we found that the equivalent width or the disc size continued to decrease (though slowly) even after $\sim$465 days of the onset of the outburst (our last optical observation). After the outburst, the optical brightness increases by $\sim$ 0.2 mag (see Figure~\ref{fig:aavso}), meaning the mass transfer happens from the central star to the inner region of the disc. The H$\alpha$ $EW$ may have recovered after the mass ejection from the star, but it decreased slowly from 59213 to 59624 except for a drastic decrease in $EW$ from 59254 to 59563, which may be due to a normal outburst around 59500. But no drastic decrease in $EW$ was found for the He line as the outer part of the disc may be only affected during the normal outburst. This slow decrease in the $EW$ may be because of less number of H$\alpha$ photons reaching the observer due to the change in the inclination angle between the disc and the line of sight. A similar decrease in the $EW$ has been reported in the BeXRB 4U~0115+63 (Figure~6 of \citealt{2001A&A...369..117N}). In the same figure, it is evident that the shape of the line profile also changed from a single-peaked to a double-peaked. This change in the viewing angle is attributed to the precession of the disc. Therefore, the observed decrease in the $EW$ in 1A~0535+262 (after the X-ray outburst) is possible due to the change in the inclination of the Be disc. This is also supported by the observed changes in the profiles of H and He lines (Figures~\ref{fig:halpha}, \ref{fig:he6678} \& \ref{fig:he7065} from MJD 59563 to 59624) from single-peaked to double-peaked and double-peaked to shell-like profiles, respectively.

The torque exerted by the Be star plays an essential role in the disc warping mechanism. The positive torque provided by the Be star through the expelling of matter always tries to keep the disc aligned with the equatorial plane of the Be star. In a torque-free case, when there is no supply of matter from the Be star to the disc, however, the disc can be easily warped towards the orbital plane of the binary. Lack of supply of matter reduces the $V$-band flux, which is originated from the inner region of the disc. Previously, the occurrence of the giant X-ray outbursts during the optical fading phase had been reported in 1A~0535+262 \citep{Camero_Arranz_2012}. \citet{Yan_2011} found an anti-correlation between the $V$-band magnitude and the $EW$ of the H$\alpha$ line prior to the 2009 outburst and interpreted this as an indication of mass loss from the inner part of the Be disc. Furthermore, we studied the $V$-band magnitude variation obtained from the AAVSO database with the Swift/BAT data (Figure~\ref{fig:aavso}).  Similar to the 2009 giant outburst (\citealt{Camero_Arranz_2012} and \citealt{Yan_2011}), we also see an increase in V-magnitude before the 2020 giant outburst. The V-magnitude was highest at $>$9.4 around MJD 58965, suggesting a mass ejection event from the Be star that led to a dimming of the continuum emission. A copious amount of this ejected material is possibly transferred to the circumnuclear disc that grows in size outward. As a result, we would expect a strong H$\alpha$ line with increasing equivalent width as seen below MJD 58920 in our study (Figure~\ref{fig:variation_halpha}). The disc at this point is significantly large (about 8$R_*$). This would lead to a strong X-ray outburst such as observed in 2020 October. After the outburst, the circumstellar disc around the Be star is truncated by the neutron star. A rapid decrease in the H$\alpha$ emission, therefore, is expected. The H$\alpha$ line had not recovered to its pre-outburst level even more than 400 days have passed since the outburst. A much faster disc recovery was seen during the 2009 outburst where the H$\alpha$ $EW$ leveled up after a year. Moreover, \citet{10.1111/j.1365-2966.2004.07743.x} and \citet{Yan_2011} found a strong He~I(7065~\AA) line during the 1994 and 2009 outbursts, suggesting the inner disc was untruncated. However, a gradual decrease in He~I(7065~\AA) strength in our observations implies that the inner region of the disc was affected during the 2020 giant outburst. 

In addition to that,  to understand the inner disc condition around the giant X-ray outburst (2020 October - December), we calculated the Full Width at Zero Intensity (FWZI) of the H$\alpha$ line. The FWZI explains the higher velocity components corresponding to the inner part of the disc. In a torque-free case, i.e., lack of supply of matter from the Be star, the density of the inner part of the disc gets reduced, thereby reducing the high-velocity component of the H$\alpha$ line (FWZI). We calculated the FWZI of the H$\alpha$ line from our pre-outburst observations and estimated its average value as $\sim$1017 km/s. However, during the peak of the outburst, the FWZI is estimated to be $\sim$771 km/s. This gives us evidence that the inner part of the decretion disc was missing prior to the giant outburst, influencing the warping of the outer part of the decretion disc and possibly causing the giant X-ray outburst from the neutron star.

\subsection{Timing and spectral properties of the pulsar with AstroSat}
During the AstroSat observation of 1A~0535+262 in the rising phase of the giant X-ray outburst, the luminosity of the pulsar was $\approx$3.91$\times$10$^{37}$ erg~s$^{-1}$ ($L/L_{\mathrm{Edd}}$ $\sim$ 0.28) in the 0.7-90.0 keV energy range.  While estimating the luminosity, we considered the distance of the source as 2.13 kpc from Gaia observation \citep{2018AJ....156...58B}. This implies that the pulsar was emitting radiation in the Sub-Eddington level (below the critical limit; see also \citealt{Sartore_2015}). The spin period of the pulsar is estimated to be 103.548 s. The average pulse profiles exhibited a simple double-peaked structure. The pulse profile of the pulsar is found to show strong energy dependence. A phase shift of $\sim$0.1 phase is observed in the profiles beyond $\sim$25 keV, indicating a marginal change in the emitting region. At hard X-ray ranges, the secondary peak gets diminished gradually making the profile single-peaked. On investigating the presence of pulsations in energy-resolved light curves, we found that 103.548 s pulsations are detected in light curves up to 110 keV, beyond which the CZTI data are background dominated. The pulse profile in soft X-ray ranges (up to 7 keV) showed  several complex dip-like features. These features are also reported in the pulse profile of the pulsar in earlier observations of 1A~0535+262 \citep{2008ApJ...672..516N,Caballero_2013} and several others, e.g., SAX~2103.5+4545 \citep{2007A&A...473..551C}, EXO~2030+375 \citep{2013ApJ...764..158N, 2015RAA....15..537N,Epili2017MNRAS.472.3455E} and GX~304-1 \citep{Jaisawal2016MNRAS.457.2749J} interpreted as due to the absorption of photons by the streams of matter locked at certain pulse phases of the neutron star. The evolution of the pulse profile from a double-peaked profile in the low energy range to a single-peaked profile in high energy range can be interpreted as due to the change in emitting region. The pulsar's pulse fraction (PF) is found to decrease initially to the lowest value in the 7-15 keV range, followed by a steady increasing trend. The high value of PF in high energy ranges indicates that most of the high energy photons contribute towards pulsation compared to the low energy photons. The lowest value of PF in 7-15 keV range can be explained as due to the diffused X-ray emission from the relatively hot plasma above the magnetic pole of the neutron star.

The pulsar spectrum in the 0.7-90 keV range was best described by a model consisting of a cut-off power-law continuum model along with two blackbody components for soft excess and a cyclotron resonance scattering feature at $\approx$46.3 keV. The observed cyclotron resonance scattering feature at $\approx$46.3 keV at a 0.7-90.0 keV luminosity of $\approx$ 3.91$\times$10$^{37}$ erg s$^{-1}$ is consistent with the earlier reported results in the rising phase of the outburst (\citealt{2019MNRAS.487L..30T,2021ApJ...917L..38K,2022MNRAS.511.1121M}). The magnetic field of the neutron star is estimated by using the relation  
\begin{equation}
    E_{CRSF}= 11.57 \times B_{12} (1+z)^{-1}
\end{equation}

Where, $E_{CRSF}$ is the cyclotron line energy in keV, $B_{12}$ is the magnetic field in units of 10$^{12}$ G, and $z$ is the gravitational redshift (z $\simeq$ 0.3 for typical neutron star). The calculated magnetic field of the pulsar is 5.2$\times$10$^{12}$ G.

\section{Conclusions}

We have studied the optical properties  of the Be/X-ray binary 1A~0535+262/HD~245770 covering the epochs before, during, and after the 2020 giant X-ray outburst over two years time span. Before the outburst, the H$\alpha$ emission lines were asymmetric and single-peaked with a broad red wing. However, during the outburst, the shape of the line profiles is opposite i.e., with a broad blue wing. After $\sim$400 days from the onset of the outburst, the profile evolved to a double peak. The equivalent width of the H$\alpha$ emission line was on an increasing trend before the giant outburst. The size of the disc derived from the H$\alpha$ $EW$ during the pre-outburst phase exceeded the 4:1 resonance radius of the decretion disc. This suggests a highly misaligned disc. The change of the line profile from a single to a double peak and highly misaligned disc suggest that the mass accretion could be from a  warped disc onto the neutron star that causes the giant outburst. We also carried out the timing and spectral analysis of the pulsar using a ToO X-ray observation with AstroSat. The X-ray pulsation at 103.55 s was detected in the light curves up to 110 keV. The pulse profile evolved from a double-peaked profile in low energy ranges to a single-peaked profile at hard X-ray ranges. In the 0.7-90.0 keV range, the source was emitting at a luminosity of 3.91$\times$10$^{37}$ erg~s$^{-1}$. The CRSF is detected at 46.3 keV, suggesting the magnetic field of the neutron star to be 5.2$\times$10$^{12}$ G. 

\section*{Acknowledgements}
We thank the anonymous reviewer for the suggestions, which helped us to improve the manuscript. The authors thank Dr. Mudit K. Srivastava and all other MFOSC-P instrument team members of Physical Research Laboratory, India, for their constant support during the optical observations at various epochs. We also thank them for providing the data reduction pipeline and other support as and when required. The authors acknowledge the entire AstroSat team of the Indian Space Research Organization (ISRO) for accepting our proposal and providing the data through the Indian Space Science Data Centre (ISSDC) platform. This work has used the data from the Soft X-ray Telescope (SXT) developed at TIFR, Mumbai, and the SXT POC at TIFR is thanked for verifying and releasing the data via the ISSDC data archive and providing the necessary software tools. This work has used the AstroSat data from the Large Area X-ray Proportional Counter (LAXPC) detectors developed at TIFR, Mumbai. We thank the LAXPC team for verifying and releasing the data via the ISSDC data archive and for providing the necessary software tools. We also extend our gratitude to the CZTI POC team members at IUCAA for helping with the augmentation of data. We are also grateful to the variable star observations community observers from the AAVSO International Database. 

\section*{DATA AVAILABILITY}
We used the optical and X-ray data from MIRO and AstroSat observatories in this work, respectively. The optical data can be shared on request. AstroSat\footnote{https://www.issdc.gov.in/astro.html}, Swift/BAT\footnote{https://swift.gsfc.nasa.gov/results/transients/1A0535p262.lc.txt}, and AAVSO\footnote{https://www.aavso.org/} data are publicly available. 


\bibliographystyle{mnras}
\bibliography{1A0535+262} 

\begin{thebibliography}{}
\makeatletter
\relax
\def\mn@urlcharsother{\let\do\@makeother \do\$\do\&\do\#\do\^\do\_\do\%\do\~}
\def\mn@doi{\begingroup\mn@urlcharsother \@ifnextchar [ {\mn@doi@}
  {\mn@doi@[]}}
\def\mn@doi@[#1]#2{\def\@tempa{#1}\ifx\@tempa\@empty \href
  {http://dx.doi.org/#2} {doi:#2}\else \href {http://dx.doi.org/#2} {#1}\fi
  \endgroup}
\def\mn@eprint#1#2{\mn@eprint@#1:#2::\@nil}
\def\mn@eprint@arXiv#1{\href {http://arxiv.org/abs/#1} {{\tt arXiv:#1}}}
\def\mn@eprint@dblp#1{\href {http://dblp.uni-trier.de/rec/bibtex/#1.xml}
  {dblp:#1}}
\def\mn@eprint@#1:#2:#3:#4\@nil{\def\@tempa {#1}\def\@tempb {#2}\def\@tempc
  {#3}\ifx \@tempc \@empty \let \@tempc \@tempb \let \@tempb \@tempa \fi \ifx
  \@tempb \@empty \def\@tempb {arXiv}\fi \@ifundefined
  {mn@eprint@\@tempb}{\@tempb:\@tempc}{\expandafter \expandafter \csname
  mn@eprint@\@tempb\endcsname \expandafter{\@tempc}}}

\bibitem[\protect\citeauthoryear{Agrawal}{Agrawal}{2006}]{agrawal2006broad}
Agrawal P.,  2006, Advances in Space Research, 38, 2989

\bibitem[\protect\citeauthoryear{Angeloni et~al.,}{Angeloni
  et~al.}{2019}]{article}
Angeloni R.,  et~al., 2019, \mn@doi [The Astronomical Journal]
  {10.3847/1538-3881/ab0cf7}, 157, 156

\bibitem[\protect\citeauthoryear{{Antia} et~al.,}{{Antia}
  et~al.}{2021}]{2021JApA...42...32A}
{Antia} H.~M.,  et~al., 2021, \mn@doi [Journal of Astrophysics and Astronomy]
  {10.1007/s12036-021-09712-8}, \href
  {https://ui.adsabs.harvard.edu/abs/2021JApA...42...32A} {42, 32}

\bibitem[\protect\citeauthoryear{{Arnaud}}{{Arnaud}}{1996}]{1996ASPC..101...17A}
{Arnaud} K.~A.,  1996, in {Jacoby} G.~H.,  {Barnes} J.,  eds,  Astronomical
  Society of the Pacific Conference Series Vol. 101, Astronomical Data Analysis
  Software and Systems V. p.~17

\bibitem[\protect\citeauthoryear{{Bailer-Jones}, {Rybizki}, {Fouesneau},
  {Mantelet}  \& {Andrae}}{{Bailer-Jones} et~al.}{2018}]{2018AJ....156...58B}
{Bailer-Jones} C.~A.~L.,  {Rybizki} J.,  {Fouesneau} M.,  {Mantelet} G.,
  {Andrae} R.,  2018, \mn@doi [\aj] {10.3847/1538-3881/aacb21}, \href
  {https://ui.adsabs.harvard.edu/abs/2018AJ....156...58B} {156, 58}

\bibitem[\protect\citeauthoryear{Bhalerao et~al.,}{Bhalerao
  et~al.}{2017}]{bhalerao2017cadmium}
Bhalerao V.,  et~al., 2017, Journal of Astrophysics and Astronomy, 38, 1

\bibitem[\protect\citeauthoryear{{Bildsten} et~al.,}{{Bildsten}
  et~al.}{1997}]{1997ApJS..113..367B}
{Bildsten} L.,  et~al., 1997, \mn@doi [\apjs] {10.1086/313060}, \href
  {https://ui.adsabs.harvard.edu/abs/1997ApJS..113..367B} {113, 367}

\bibitem[\protect\citeauthoryear{Caballero et~al.,}{Caballero
  et~al.}{2013}]{Caballero_2013}
Caballero I.,  et~al., 2013, \mn@doi [The Astrophysical Journal]
  {10.1088/2041-8205/764/2/l23}, 764, L23

\bibitem[\protect\citeauthoryear{{Camero Arranz}, {Wilson}, {Finger}  \&
  {Reglero}}{{Camero Arranz} et~al.}{2007}]{2007A&A...473..551C}
{Camero Arranz} A.,  {Wilson} C.~A.,  {Finger} M.~H.,   {Reglero} V.,  2007,
  \mn@doi [\aap] {10.1051/0004-6361:20077398}, \href
  {https://ui.adsabs.harvard.edu/abs/2007A&A...473..551C} {473, 551}

\bibitem[\protect\citeauthoryear{Camero-Arranz et~al.,}{Camero-Arranz
  et~al.}{2012}]{Camero_Arranz_2012}
Camero-Arranz A.,  et~al., 2012, \mn@doi [The Astrophysical Journal]
  {10.1088/0004-637x/754/1/20}, 754, 20

\bibitem[\protect\citeauthoryear{{Clark} et~al.,}{{Clark}
  et~al.}{1998}]{1998MNRAS.294..165C}
{Clark} J.~S.,  et~al., 1998, \mn@doi [\mnras]
  {10.1046/j.1365-8711.1998.01257.x}, \href
  {https://ui.adsabs.harvard.edu/abs/1998MNRAS.294..165C} {294, 165}

\bibitem[\protect\citeauthoryear{Clark et~al.,}{Clark
  et~al.}{1999}]{10.1046/j.1365-8711.1999.02112.x}
Clark J.~S.,  et~al., 1999, \mn@doi [Monthly Notices of the Royal Astronomical
  Society] {10.1046/j.1365-8711.1999.02112.x}, 302, 167

\bibitem[\protect\citeauthoryear{Coe, Reig, McBride, Galache  \& Fabregat}{Coe
  et~al.}{2006}]{10.1111/j.1365-2966.2006.10127.x}
Coe M.~J.,  Reig P.,  McBride V.~A.,  Galache J.~L.,   Fabregat J.,  2006,
  \mn@doi [Monthly Notices of the Royal Astronomical Society]
  {10.1111/j.1365-2966.2006.10127.x}, 368, 447

\bibitem[\protect\citeauthoryear{{Epili}, {Naik}, {Jaisawal}  \&
  {Gupta}}{{Epili} et~al.}{2017}]{Epili2017MNRAS.472.3455E}
{Epili} P.,  {Naik} S.,  {Jaisawal} G.~K.,   {Gupta} S.,  2017, \mn@doi
  [\mnras] {10.1093/mnras/stx2247}, \href
  {https://ui.adsabs.harvard.edu/abs/2017MNRAS.472.3455E} {472, 3455}

\bibitem[\protect\citeauthoryear{Finger, Cominsky, Wilson, Harmon  \&
  Fishman}{Finger et~al.}{1994}]{finger1994hard}
Finger M.~H.,  Cominsky L.~R.,  Wilson R.~B.,  Harmon B.~A.,   Fishman G.~J.,
  1994, in AIP Conference Proceedings. pp 459--462

\bibitem[\protect\citeauthoryear{Giangrande, Giovannelli, Bartolini, Guarnieri
  \& Piccioni}{Giangrande et~al.}{1980}]{giangrande1980optical}
Giangrande A.,  Giovannelli F.,  Bartolini C.,  Guarnieri A.,   Piccioni A.,
  1980, Astronomy and Astrophysics Supplement Series, 40, 289

\bibitem[\protect\citeauthoryear{Grundstrom et~al.,}{Grundstrom
  et~al.}{2007}]{Grundstrom_2007}
Grundstrom E.~D.,  et~al., 2007, \mn@doi [The Astrophysical Journal]
  {10.1086/510508}, 656, 431

\bibitem[\protect\citeauthoryear{Haigh, Coe, Steele  \& Fabregat}{Haigh
  et~al.}{1999}]{10.1046/j.1365-8711.1999.03148.x}
Haigh N.~J.,  Coe M.~J.,  Steele I.~A.,   Fabregat J.,  1999, \mn@doi [Monthly
  Notices of the Royal Astronomical Society]
  {10.1046/j.1365-8711.1999.03148.x}, 310, L21

\bibitem[\protect\citeauthoryear{Haigh, Coe  \& Fabregat}{Haigh
  et~al.}{2004}]{10.1111/j.1365-2966.2004.07743.x}
Haigh N.~J.,  Coe M.~J.,   Fabregat J.,  2004, \mn@doi [Monthly Notices of the
  Royal Astronomical Society] {10.1111/j.1365-2966.2004.07743.x}, 350, 1457

\bibitem[\protect\citeauthoryear{{Hanuschik}}{{Hanuschik}}{1989}]{1989Ap&SS.161...61H}
{Hanuschik} R.~W.,  1989, \mn@doi [\apss] {10.1007/BF00653238}, \href
  {https://ui.adsabs.harvard.edu/abs/1989Ap&SS.161...61H} {161, 61}

\bibitem[\protect\citeauthoryear{{Herbig}}{{Herbig}}{1995}]{1995ARA&A..33...19H}
{Herbig} G.~H.,  1995, \mn@doi [\araa] {10.1146/annurev.aa.33.090195.000315},
  \href {https://ui.adsabs.harvard.edu/abs/1995ARA&A..33...19H} {33, 19}

\bibitem[\protect\citeauthoryear{{Huang}}{{Huang}}{1972}]{1972ApJ...171..549H}
{Huang} S.-S.,  1972, \mn@doi [\apj] {10.1086/151309}, \href
  {https://ui.adsabs.harvard.edu/abs/1972ApJ...171..549H} {171, 549}

\bibitem[\protect\citeauthoryear{{Hummel}}{{Hummel}}{1998}]{1998A&A...330..243H}
{Hummel} W.,  1998, \aap, \href
  {https://ui.adsabs.harvard.edu/abs/1998A&A...330..243H} {330, 243}

\bibitem[\protect\citeauthoryear{{Jaisawal}, {Naik}  \& {Epili}}{{Jaisawal}
  et~al.}{2016}]{Jaisawal2016MNRAS.457.2749J}
{Jaisawal} G.~K.,  {Naik} S.,   {Epili} P.,  2016, \mn@doi [\mnras]
  {10.1093/mnras/stw085}, \href
  {https://ui.adsabs.harvard.edu/abs/2016MNRAS.457.2749J} {457, 2749}

\bibitem[\protect\citeauthoryear{{Jaisawal} et~al.,}{{Jaisawal}
  et~al.}{2020a}]{Jaisawal2020ATel14179....1J}
{Jaisawal} G.~K.,  et~al., 2020a, The Astronomer's Telegram, \href
  {https://ui.adsabs.harvard.edu/abs/2020ATel14179....1J} {14179, 1}

\bibitem[\protect\citeauthoryear{{Jaisawal} et~al.,}{{Jaisawal}
  et~al.}{2020b}]{Jaisawal2020ATel14227....1J}
{Jaisawal} G.~K.,  et~al., 2020b, The Astronomer's Telegram, \href
  {https://ui.adsabs.harvard.edu/abs/2020ATel14227....1J} {14227, 1}

\bibitem[\protect\citeauthoryear{{Kendziorra} et~al.,}{{Kendziorra}
  et~al.}{1994}]{1994A&A...291L..31K}
{Kendziorra} E.,  et~al., 1994, \aap, \href
  {https://ui.adsabs.harvard.edu/abs/1994A&A...291L..31K} {291, L31}

\bibitem[\protect\citeauthoryear{{Kong} et~al.,}{{Kong}
  et~al.}{2021}]{2021ApJ...917L..38K}
{Kong} L.~D.,  et~al., 2021, \mn@doi [\apjl] {10.3847/2041-8213/ac1ad3}, \href
  {https://ui.adsabs.harvard.edu/abs/2021ApJ...917L..38K} {917, L38}

\bibitem[\protect\citeauthoryear{{Krimm} et~al.,}{{Krimm}
  et~al.}{2013}]{Krimm2013ApJS..209...14K}
{Krimm} H.~A.,  et~al., 2013, \mn@doi [\apjs] {10.1088/0067-0049/209/1/14},
  \href {https://ui.adsabs.harvard.edu/abs/2013ApJS..209...14K} {209, 14}

\bibitem[\protect\citeauthoryear{{Kumar} et~al.,}{{Kumar}
  et~al.}{2022}]{2022MNRAS.510.4265K}
{Kumar} V.,  et~al., 2022, \mn@doi [\mnras] {10.1093/mnras/stab3772}, \href
  {https://ui.adsabs.harvard.edu/abs/2022MNRAS.510.4265K} {510, 4265}

\bibitem[\protect\citeauthoryear{{Leahy}}{{Leahy}}{1987}]{1987A&A...180..275L}
{Leahy} D.~A.,  1987, \aap, \href
  {https://ui.adsabs.harvard.edu/abs/1987A&A...180..275L} {180, 275}

\bibitem[\protect\citeauthoryear{{Malacaria}, {Kollatschny}, {Whelan},
  {Santangelo}, {Klochkov}, {McBride}  \& {Ducci}}{{Malacaria}
  et~al.}{2017}]{2017A&A...603A..24M}
{Malacaria} C.,  {Kollatschny} W.,  {Whelan} E.,  {Santangelo} A.,  {Klochkov}
  D.,  {McBride} V.,   {Ducci} L.,  2017, \mn@doi [\aap]
  {10.1051/0004-6361/201730538}, \href
  {https://ui.adsabs.harvard.edu/abs/2017A&A...603A..24M} {603, A24}

\bibitem[\protect\citeauthoryear{{Mandal} \& {Pal}}{{Mandal} \&
  {Pal}}{2022}]{2022MNRAS.511.1121M}
{Mandal} M.,  {Pal} S.,  2022, \mn@doi [\mnras] {10.1093/mnras/stac111}, \href
  {https://ui.adsabs.harvard.edu/abs/2022MNRAS.511.1121M} {511, 1121}

\bibitem[\protect\citeauthoryear{{Mandal}, {Pal}, {Hazra}, {Jana}, {Bhunia}  \&
  {Ghanta}}{{Mandal} et~al.}{2020}]{2020ATel14157....1M}
{Mandal} M.,  {Pal} S.,  {Hazra} M.,  {Jana} A.,  {Bhunia} B.,   {Ghanta} A.,
  2020, The Astronomer's Telegram, \href
  {https://ui.adsabs.harvard.edu/abs/2020ATel14157....1M} {14157, 1}

\bibitem[\protect\citeauthoryear{Martin, Pringle, Tout  \& Lubow}{Martin
  et~al.}{2011}]{10.1111/j.1365-2966.2011.19231.x}
Martin R.~G.,  Pringle J.~E.,  Tout C.~A.,   Lubow S.~H.,  2011, \mn@doi
  [Monthly Notices of the Royal Astronomical Society]
  {10.1111/j.1365-2966.2011.19231.x}, 416, 2827

\bibitem[\protect\citeauthoryear{Martin, Nixon, Armitage, Lubow  \&
  Price}{Martin et~al.}{2014}]{Martin_2014_a}
Martin R.~G.,  Nixon C.,  Armitage P.~J.,  Lubow S.~H.,   Price D.~J.,  2014,
  \mn@doi [The Astrophysical Journal] {10.1088/2041-8205/790/2/l34}, 790, L34

\bibitem[\protect\citeauthoryear{Monageng, McBride, Coe, Steele  \&
  Reig}{Monageng et~al.}{2016}]{10.1093/mnras/stw2354}
Monageng I.~M.,  McBride V.~A.,  Coe M.~J.,  Steele I.~A.,   Reig P.,  2016,
  \mn@doi [Monthly Notices of the Royal Astronomical Society]
  {10.1093/mnras/stw2354}, 464, 572

\bibitem[\protect\citeauthoryear{Moritani et~al.,}{Moritani
  et~al.}{2010}]{10.1111/j.1365-2966.2010.16454.x}
Moritani Y.,  et~al., 2010, \mn@doi [Monthly Notices of the Royal Astronomical
  Society] {10.1111/j.1365-2966.2010.16454.x}, 405, 467

\bibitem[\protect\citeauthoryear{Moritani, Nogami, Okazaki, Imada, Kambe,
  Honda, Hashimoto  \& Ichikawa}{Moritani et~al.}{2011}]{10.1093/pasj/63.4.L25}
Moritani Y.,  Nogami D.,  Okazaki A.~T.,  Imada A.,  Kambe E.,  Honda S.,
  Hashimoto O.,   Ichikawa K.,  2011, \mn@doi [Publications of the Astronomical
  Society of Japan] {10.1093/pasj/63.4.L25}, 63, L25

\bibitem[\protect\citeauthoryear{Moritani et~al.,}{Moritani
  et~al.}{2013}]{article_Moritani}
Moritani Y.,  et~al., 2013, \mn@doi [Publications of the Astronomical Society
  of Japan] {10.1093/pasj/65.4.83}, 65

\bibitem[\protect\citeauthoryear{{Naik} \& {Jaisawal}}{{Naik} \&
  {Jaisawal}}{2015}]{2015RAA....15..537N}
{Naik} S.,  {Jaisawal} G.~K.,  2015, \mn@doi [Research in Astronomy and
  Astrophysics] {10.1088/1674-4527/15/4/007}, \href
  {https://ui.adsabs.harvard.edu/abs/2015RAA....15..537N} {15, 537}

\bibitem[\protect\citeauthoryear{{Naik} et~al.,}{{Naik}
  et~al.}{2008}]{2008ApJ...672..516N}
{Naik} S.,  et~al., 2008, \mn@doi [\apj] {10.1086/523295}, \href
  {https://ui.adsabs.harvard.edu/abs/2008ApJ...672..516N} {672, 516}

\bibitem[\protect\citeauthoryear{{Naik}, {Maitra}, {Jaisawal}  \&
  {Paul}}{{Naik} et~al.}{2013}]{2013ApJ...764..158N}
{Naik} S.,  {Maitra} C.,  {Jaisawal} G.~K.,   {Paul} B.,  2013, \mn@doi [\apj]
  {10.1088/0004-637X/764/2/158}, \href
  {https://ui.adsabs.harvard.edu/abs/2013ApJ...764..158N} {764, 158}

\bibitem[\protect\citeauthoryear{{Negueruela} \& {Okazaki}}{{Negueruela} \&
  {Okazaki}}{2001}]{2001A&A...369..108N}
{Negueruela} I.,  {Okazaki} A.~T.,  2001, \mn@doi [\aap]
  {10.1051/0004-6361:20010146}, \href
  {https://ui.adsabs.harvard.edu/abs/2001A&A...369..108N} {369, 108}

\bibitem[\protect\citeauthoryear{{Negueruela}, {Okazaki}, {Fabregat}, {Coe},
  {Munari}  \& {Tomov}}{{Negueruela} et~al.}{2001}]{2001A&A...369..117N}
{Negueruela} I.,  {Okazaki} A.~T.,  {Fabregat} J.,  {Coe} M.~J.,  {Munari} U.,
   {Tomov} T.,  2001, \mn@doi [\aap] {10.1051/0004-6361:20010077}, \href
  {https://ui.adsabs.harvard.edu/abs/2001A&A...369..117N} {369, 117}

\bibitem[\protect\citeauthoryear{Okazaki \& Hayasaki}{Okazaki \&
  Hayasaki}{2007}]{okazaki2007interaction}
Okazaki A.,  Hayasaki K.,  2007, in Active OB-Stars: Laboratories for Stellare
  and Circumstellar Physics. p.~395

\bibitem[\protect\citeauthoryear{Okazaki \& Negueruela}{Okazaki \&
  Negueruela}{2001}]{okazaki2001natural}
Okazaki A.,  Negueruela I.,  2001, Astronomy \& Astrophysics, 377, 161

\bibitem[\protect\citeauthoryear{Okazaki, Bate, Ogilvie  \& Pringle}{Okazaki
  et~al.}{2002}]{10.1046/j.1365-8711.2002.05960.x}
Okazaki A.~T.,  Bate M.~R.,  Ogilvie G.~I.,   Pringle J.~E.,  2002, \mn@doi
  [Monthly Notices of the Royal Astronomical Society]
  {10.1046/j.1365-8711.2002.05960.x}, 337, 967

\bibitem[\protect\citeauthoryear{Okazaki, Hayasaki  \& Moritani}{Okazaki
  et~al.}{2013}]{10.1093/pasj/65.2.41}
Okazaki A.~T.,  Hayasaki K.,   Moritani Y.,  2013, \mn@doi [Publications of the
  Astronomical Society of Japan] {10.1093/pasj/65.2.41}, 65

\bibitem[\protect\citeauthoryear{{Parmar}, {White}  \& {Stella}}{{Parmar}
  et~al.}{1989}]{1989ApJ...338..373P}
{Parmar} A.~N.,  {White} N.~E.,   {Stella} L.,  1989, \mn@doi [\apj]
  {10.1086/167205}, \href
  {https://ui.adsabs.harvard.edu/abs/1989ApJ...338..373P} {338, 373}

\bibitem[\protect\citeauthoryear{Paul \& Naik}{Paul \&
  Naik}{2011}]{paul2011transient}
Paul B.,  Naik S.,  2011, Bull. Astr. Soc. India, 39, 1

\bibitem[\protect\citeauthoryear{{Porter}}{{Porter}}{1999}]{1999A&A...348..512P}
{Porter} J.~M.,  1999, \aap, \href
  {https://ui.adsabs.harvard.edu/abs/1999A&A...348..512P} {348, 512}

\bibitem[\protect\citeauthoryear{Porter \& Rivinius}{Porter \&
  Rivinius}{2003}]{porter2003classical}
Porter J.~M.,  Rivinius T.,  2003, Publications of the Astronomical Society of
  the Pacific, 115, 1153

\bibitem[\protect\citeauthoryear{Ramadevi et~al.,}{Ramadevi
  et~al.}{2017}]{ramadevi2017early}
Ramadevi M.,  et~al., 2017, Journal of Astrophysics and Astronomy, 38, 1

\bibitem[\protect\citeauthoryear{Reig}{Reig}{2011}]{reig2011x}
Reig P.,  2011, Astrophysics and Space Science, 332, 1

\bibitem[\protect\citeauthoryear{{Reig} \& {Blinov}}{{Reig} \&
  {Blinov}}{2018}]{2018A&A...619A..19R}
{Reig} P.,  {Blinov} D.,  2018, \mn@doi [\aap] {10.1051/0004-6361/201833649},
  \href {https://ui.adsabs.harvard.edu/abs/2018A&A...619A..19R} {619, A19}

\bibitem[\protect\citeauthoryear{{Reig} \& {Roche}}{{Reig} \&
  {Roche}}{1999}]{1999MNRAS.306..100R}
{Reig} P.,  {Roche} P.,  1999, \mn@doi [\mnras]
  {10.1046/j.1365-8711.1999.02473.x}, \href
  {https://ui.adsabs.harvard.edu/abs/1999MNRAS.306..100R} {306, 100}

\bibitem[\protect\citeauthoryear{{Reig} \& {Zezas}}{{Reig} \&
  {Zezas}}{2014}]{2014A&A...561A.137R}
{Reig} P.,  {Zezas} A.,  2014, \mn@doi [\aap] {10.1051/0004-6361/201321408},
  \href {https://ui.adsabs.harvard.edu/abs/2014A&A...561A.137R} {561, A137}

\bibitem[\protect\citeauthoryear{{Reig, P.} \& {Nespoli, E.}}{{Reig, P.} \&
  {Nespoli, E.}}{2013}]{refId013}
{Reig, P.} {Nespoli, E.} 2013, \mn@doi [A\&A] {10.1051/0004-6361/201219806},
  551, A1

\bibitem[\protect\citeauthoryear{{Reig}, {Negueruela}, {Fabregat}, {Chato}  \&
  {Coe}}{{Reig} et~al.}{2005}]{Reig2005A&A...440.1079R}
{Reig} P.,  {Negueruela} I.,  {Fabregat} J.,  {Chato} R.,   {Coe} M.~J.,  2005,
  \mn@doi [\aap] {10.1051/0004-6361:20053124}, \href
  {https://ui.adsabs.harvard.edu/abs/2005A&A...440.1079R} {440, 1079}

\bibitem[\protect\citeauthoryear{{Reig}, {Larionov}, {Negueruela}, {Arkharov}
  \& {Kudryavtseva}}{{Reig} et~al.}{2007}]{2007A&A...462.1081R}
{Reig} P.,  {Larionov} V.,  {Negueruela} I.,  {Arkharov} A.~A.,
  {Kudryavtseva} N.~A.,  2007, \mn@doi [\aap] {10.1051/0004-6361:20066217},
  \href {https://ui.adsabs.harvard.edu/abs/2007A&A...462.1081R} {462, 1081}

\bibitem[\protect\citeauthoryear{Rosenberg, Eyles, Skinner  \&
  Willmore}{Rosenberg et~al.}{1975}]{rosenberg1975observations}
Rosenberg F.,  Eyles C.,  Skinner G.,   Willmore A.,  1975, Nature, 256, 628

\bibitem[\protect\citeauthoryear{Sartore, Jourdain  \& Roques}{Sartore
  et~al.}{2015}]{Sartore_2015}
Sartore N.,  Jourdain E.,   Roques J.~P.,  2015, \mn@doi [The Astrophysical
  Journal] {10.1088/0004-637x/806/2/193}, 806, 193

\bibitem[\protect\citeauthoryear{{Singh} et~al.,}{{Singh}
  et~al.}{2017}]{2017JApA...38...29S}
{Singh} K.~P.,  et~al., 2017, \mn@doi [Journal of Astrophysics and Astronomy]
  {10.1007/s12036-017-9448-7}, \href
  {https://ui.adsabs.harvard.edu/abs/2017JApA...38...29S} {38, 29}

\bibitem[\protect\citeauthoryear{{Srivastava}, {Jangra}, {Dixit}, {Munjal},
  {Arora}  \& {Mavani}}{{Srivastava} et~al.}{2018}]{2018SPIE10702E..4IS}
{Srivastava} M.~K.,  {Jangra} M.,  {Dixit} V.,  {Munjal} B.~S.,  {Arora} H.,
  {Mavani} T.,  2018, in {Evans} C.~J.,  {Simard} L.,   {Takami} H.,  eds,
  Society of Photo-Optical Instrumentation Engineers (SPIE) Conference Series
  Vol. 10702, Ground-based and Airborne Instrumentation for Astronomy VII. p.
  107024I, \mn@doi{10.1117/12.2309306}

\bibitem[\protect\citeauthoryear{Srivastava, Kumar, Dixit, Patel, Jangra,
  Rajpurohit  \& Mathur}{Srivastava et~al.}{2021}]{srivastava2021design}
Srivastava M.~K.,  Kumar V.,  Dixit V.,  Patel A.,  Jangra M.,  Rajpurohit A.,
   Mathur S.,  2021, Experimental Astronomy, 51, 345

\bibitem[\protect\citeauthoryear{{Tandon} et~al.,}{{Tandon}
  et~al.}{2017}]{2017AJ....154..128T}
{Tandon} S.~N.,  et~al., 2017, \mn@doi [\aj] {10.3847/1538-3881/aa8451}, \href
  {https://ui.adsabs.harvard.edu/abs/2017AJ....154..128T} {154, 128}

\bibitem[\protect\citeauthoryear{{Terada} et~al.,}{{Terada}
  et~al.}{2006}]{2006ApJ...648L.139T}
{Terada} Y.,  et~al., 2006, \mn@doi [\apjl] {10.1086/508018}, \href
  {https://ui.adsabs.harvard.edu/abs/2006ApJ...648L.139T} {648, L139}

\bibitem[\protect\citeauthoryear{{Tsygankov}, {Doroshenko}, {Mushtukov},
  {Suleimanov}, {Lutovinov}  \& {Poutanen}}{{Tsygankov}
  et~al.}{2019}]{2019MNRAS.487L..30T}
{Tsygankov} S.~S.,  {Doroshenko} V.,  {Mushtukov} A.~A.,  {Suleimanov} V.~F.,
  {Lutovinov} A.~A.,   {Poutanen} J.,  2019, \mn@doi [\mnras]
  {10.1093/mnrasl/slz079}, \href
  {https://ui.adsabs.harvard.edu/abs/2019MNRAS.487L..30T} {487, L30}

\bibitem[\protect\citeauthoryear{{Verner}, {Ferland}, {Korista}  \&
  {Yakovlev}}{{Verner} et~al.}{1996}]{1996ApJ...465..487V}
{Verner} D.~A.,  {Ferland} G.~J.,  {Korista} K.~T.,   {Yakovlev} D.~G.,  1996,
  \mn@doi [\apj] {10.1086/177435}, \href
  {https://ui.adsabs.harvard.edu/abs/1996ApJ...465..487V} {465, 487}

\bibitem[\protect\citeauthoryear{{Wilms}, {Allen}  \& {McCray}}{{Wilms}
  et~al.}{2000}]{2000ApJ...542..914W}
{Wilms} J.,  {Allen} A.,   {McCray} R.,  2000, \mn@doi [\apj] {10.1086/317016},
  \href {https://ui.adsabs.harvard.edu/abs/2000ApJ...542..914W} {542, 914}

\bibitem[\protect\citeauthoryear{Yadav et~al.,}{Yadav
  et~al.}{2016}]{yadav2016large}
Yadav J.,  et~al., 2016, in Space Telescopes and Instrumentation 2016:
  Ultraviolet to Gamma Ray. pp 374--388

\bibitem[\protect\citeauthoryear{Yan, Li  \& Liu}{Yan et~al.}{2011}]{Yan_2011}
Yan J.,  Li H.,   Liu Q.,  2011, \mn@doi [The Astrophysical Journal]
  {10.1088/0004-637x/744/1/37}, 744, 37

\bibitem[\protect\citeauthoryear{{Zamanov}, {Stoyanov}, {Mart{\'\i}}, {Tomov},
  {Belcheva}, {Luque-Escamilla}  \& {Latev}}{{Zamanov}
  et~al.}{2013}]{Zamanov2013A&A...559A..87Z}
{Zamanov} R.,  {Stoyanov} K.,  {Mart{\'\i}} J.,  {Tomov} N.~A.,  {Belcheva} G.,
   {Luque-Escamilla} P.~L.,   {Latev} G.,  2013, \mn@doi [\aap]
  {10.1051/0004-6361/201321991}, \href
  {https://ui.adsabs.harvard.edu/abs/2013A&A...559A..87Z} {559, A87}

\bibitem[\protect\citeauthoryear{Ziolkowski}{Ziolkowski}{2002}]{ziolkowski2002x}
Ziolkowski J.,  2002, Memorie della Societa Astronomica Italiana, 73, 1038

\makeatother
\end{thebibliography}




\bsp	
\label{lastpage}
\end{document}